\documentclass[numberedappendix]{emulateapj}

\newcommand{\thetabold}{\mbox{\boldmath$\theta$}}
\newcommand{\sigmabold}{\mbox{\boldmath$\sigma$}}
\newcommand{\B}{\textsc{Bayes-ME}}

\begin{document}

\title{Evidence for Quasi-isotropic Magnetic Fields from \\ \emph{Hinode} Quiet Sun Observations}
\email{aasensio@iac.es}
%
\author{A. Asensio Ramos}
\affil{Instituto de Astrof\'{\i}sica de Canarias, 38205, La Laguna, Tenerife, Spain}


\begin{abstract}
Some recent investigations of spectropolarimetric observations of
the Zeeman effect in the Fe \textsc{i} lines at 630 nm
carried out with the \emph{Hinode} solar space telescope have
concluded that the strength of the magnetic field vector in the internetwork regions of
the quiet Sun is in the hG regime and that its inclination is predominantly horizontal.
We critically reconsider the analysis of such observations and carry out a complete
Bayesian analysis with the aim of extracting as much information as possible from them, 
including error bars.
We apply the recently developed \B\ code that carries out a complete
Bayesian inference for Milne-Eddington atmospheres. The sampling of the posterior
distribution function is obtained with a Markov Chain Monte Carlo scheme and the
marginal distributions are analyzed in detail. The Kullback-Leibler divergence is
used to study the extent to which the observations introduce new information in the 
inference process resulting in sufficiently constrained parameters.
Our analysis clearly shows that only upper limits to the magnetic field strength 
can be inferred with fields in the kG regime completely discarded. Furthermore,
the noise level present in the analyzed \emph{Hinode} observations induces
a substantial loss of information for constraining the azimuth of the magnetic
field. Concerning the inclination of the field, we demonstrate that some information is
available to constrain it for those pixels with the largest polarimetric signal. The results also
point out that the field in pixels with small polarimetric signals can be nicely
reproduced in terms of a quasi-isotropic distribution.
\end{abstract}
\keywords{magnetic fields --- Sun: atmosphere -- Sun: magnetic fields --- line: profiles --- polarization}



\section{Introduction}
Particularly during the last decade, we have witnessed an increasing interest in the
investigation of the magnetism of the quiet solar photosphere. There are at least two reasons for
this. On the one hand, improvements in the sensitivity of 
spectro-polarimeters have allowed us to carry out spectropolarimetric 
observations of great quality in regions of very low magnetic flux. On the
other hand, since the quiet Sun covers a fraction larger than 90\% of the
solar surface, it is important to investigate whether the magnetism of the quiet
photosphere plays a significant role on the heating of the outer regions.
In case the magnetic energy stored in these apparently
non-magnetic zones turns out to be substantial, as claimed by \cite{trujillo_nature04},
its impact on the overall energy balance of the 
solar atmosphere should be accounted for.

\begin{figure*}[!t]
\centering
\plottwo{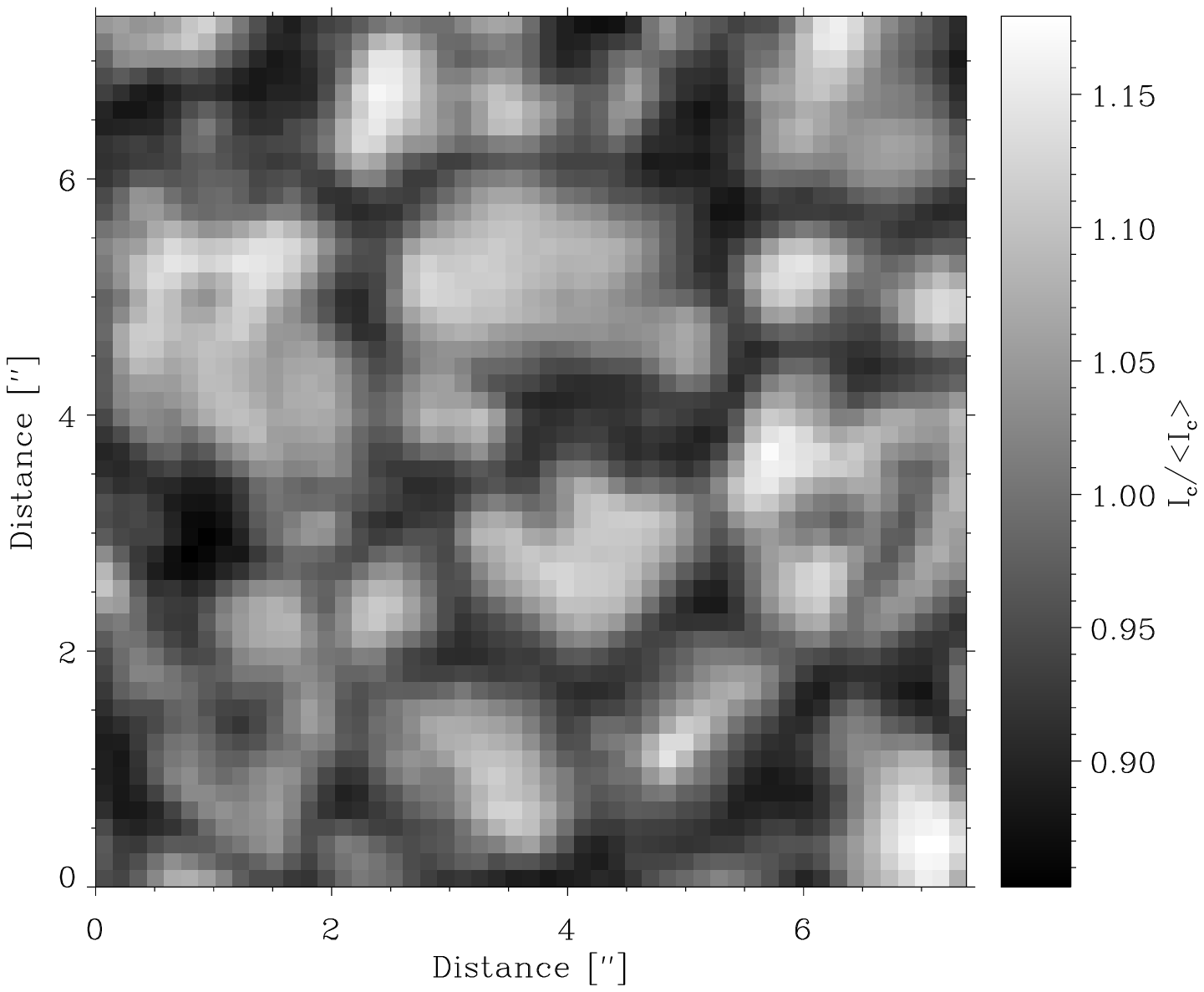}{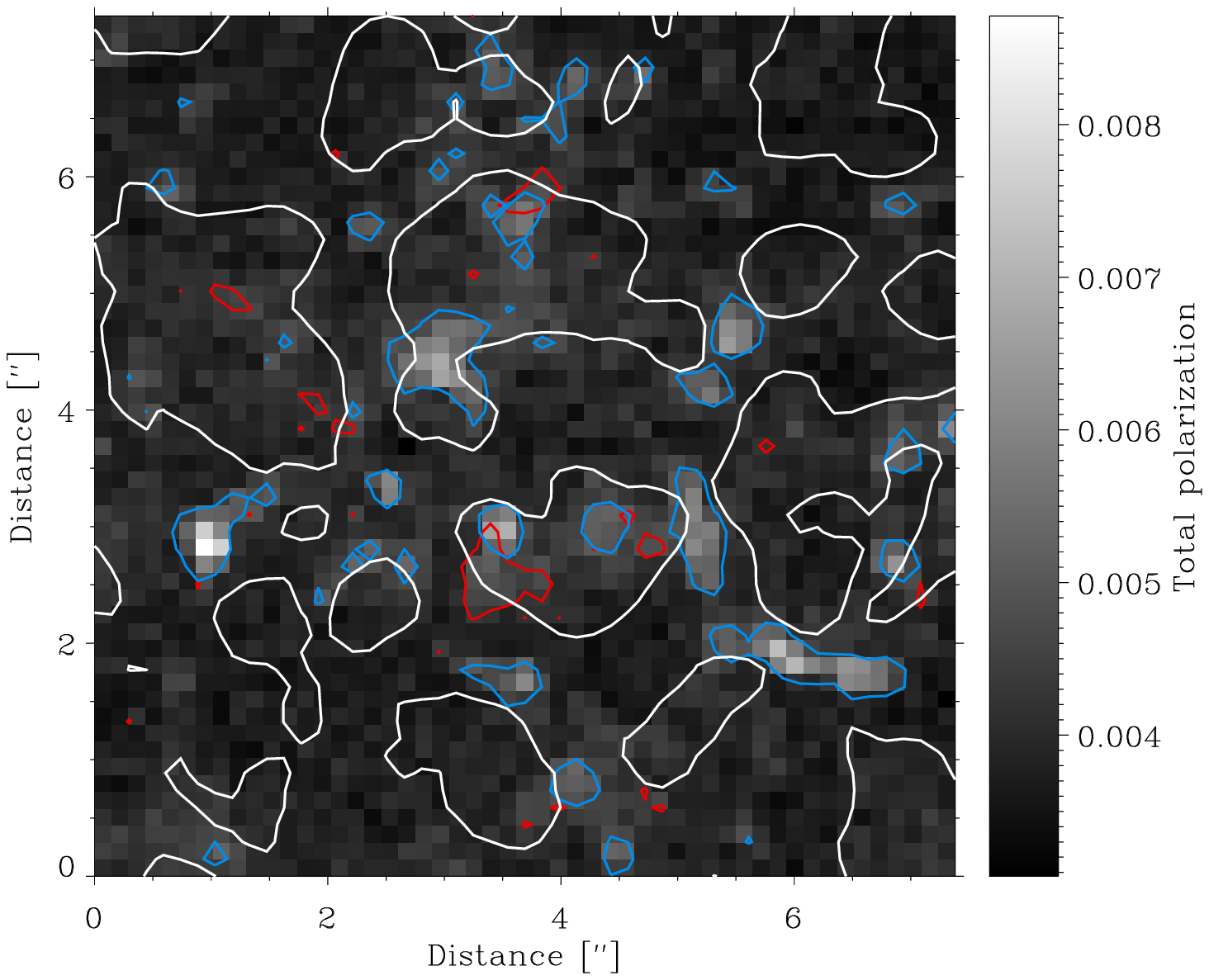}
\caption{The left panel shows the observed intensity (normalized to the average intensity) in the
continuum surrounding the 630 nm Fe \textsc{i} doublet. The background image on the right panel presents the total
polarization normalized to the average intensity, $\int(Q^2+U^2+V^2)^{1/2} d\lambda/ \langle I \rangle$.
The red and blue contours delineate regions of $\int(Q^2+U^2)^{1/2} d\lambda/ \langle I \rangle > 0.0035$ 
and $\int(V^2)^{1/2} d\lambda/ \langle I \rangle > 0.003$, respectively.
The white contours indicate regions where $I_c/\langle I_c \rangle=1.02$.}
\label{fig:maps}
\end{figure*}

The Zeeman polarization signals observed in the quiet Sun are very weak and, even with the
best spectropolarimeters, it turns out difficult to detect them 
and to interpret reliably the observations. One of the main causes for such weak polarization signals is 
that the filling factor of the magnetic field in the resolution elements of ground based
telescopes is very small, of the order of 1\% \citep[e.g.,][]{stenflo94,lin95,dominguez03,khomenko03,martinez_gonzalez06}.
The reason for this small filling factor may lie in the fact that the field is organized at
very small scales and Stokes $V$ Zeeman signals cancel effectively when averaging fields with several
inclinations in the resolution element (linear polarization also cancels when the azimuth
is also random). Understanding the probability distribution of field 
strengths has produced a long debate that is not still resolved 
\citep{lin95,lin_rimmele99,socasnavarro02,dominguez03,khomenko03,socas_pillet_lites04,lites_socas04,martinez_gonzalez_spw4_06,martinez_gonzalez06,dominguez06,orozco_hinode07,marian08,ramirez_velez08}.
The reason is that diagnostic tools based on
the Zeeman effect suffer from cancellation effects and it
is not clear to what extent the probability distribution inferred from the residual Zeeman 
signals can be compared to the probability distribution of the field strength inside the 
pixel. Therefore, it is important to develop and apply alternative diagnostic tools that do not
suffer from such cancellations. In this respect, we note the
efforts of \cite{trujillo_nature04,trujillo_asensio_shchukina_spw4_06} applying techniques 
based on the Hanle effect to infer properties about the magnetic field in the quiet Sun. They concluded that there is 
a substantial amount of ``hidden'' magnetic energy in the internetwork regions 
of the quiet photosphere. Most of these ``hidden'' fields remain undetected to the Zeeman effect because 
they are tangled at scales below the resolution of present telescopes or because they are
inherently weak. Others have applied 
Zeeman-based techniques in lines with strong hyperfine perturbations, which make them
sensitive to the strength of the magnetic field 
\citep{arturo02,arturo06,ramirez07,asensio_mn07,ramirez_velez08}.
The results go in the direction of ubiquitous fields with strengths in the hG regime, which
supports the Hanle-effect conclusion of \cite{trujillo_nature04,trujillo_asensio_shchukina_spw4_06}.

Recent results \citep{marian_clv08} indicate that there is not any apparent variation of the 
polarimetric properties of the Fe \textsc{i} near-IR doublet at 1.56 $\mu$m with the
heliocentric angle. This suggests that the magnetism of the quiet Sun remains approximately
the same for every observing angle, thus giving weight to the fact that the distribution of
magnetic field vectors is quasi-isotropic for the resolutions of ground based telescopes and
discards a network-like scenario for the internetwork magnetism. Based on these observations,
\cite{arturo_sf2a08} suggest a scenario in which the observed Zeeman signals are a result
of statistical fluctuations of a quasi-isotropic vector field. This scenario suggests a way to
reconcile simultaneously the Zeeman and the Hanle results. It also gives an explanation
to the reason why the unsigned magnetic flux seems to be the same regardless of the
spatial resolution of the observation.

Recent observations carried out with the spectropolarimeter \citep[SP;][]{lites_hinode01} 
aboard \emph{Hinode} \citep{kosugi_hinode07} have led to the detection of large amounts of
linear polarization signals in the Fe \textsc{i} lines at 630 nm at unprecedent spatial resolutions of $\sim$0.3$"$. The 
interpretation of these signals based on simple inversion methods \citep{lites08} 
or using more elaborate Milne-Eddington models \citep{orozco_hinode07,lites08} suggest the presence of 
magnetic fields with a large horizontal component. The probability distribution
of field inclinations given by \cite{orozco_hinode07} suggests that the majority of
fields found in the field-of-view are horizontal (with inclinations $\gamma=90^\circ \pm 15^\circ$), 
with a certain (small) fraction of fields being almost vertical. It is also necessary
to point out that \cite{orozco_mhd07} argued, with the aid of numerical
magneto-hydrodynamical simulations, that it is possible to get a good estimation of 
the magnetic field strength from Milne-Eddington inversions of Hinode data.

If the apparent lack of variation of the polarimetric properties with the heliocentric
angle \citep{marian_clv08} and the apparent large presence of horizontal fields at disk
center \citep{orozco_hinode07,lites08} are put together, this would mean that,
irrespective of the location in the solar disk, the fields would always be predominantly 
horizontal. This result indicates that the distribution of magnetic field vectors is indeed 
close to isotropic since an isotropic vector field distribution always presents a much 
larger amount of horizontal than vertical field (for all external observers). Therefore, it 
is of great interest to verify to what extent are the observations (with the presence of noise and the
ensuing uncertainties) unambiguously pointing to predominantly horizontal fields.

To this end, we apply \B, the Bayesian inference code for Milne-Eddington
atmospheres developed by \cite{asensio_martinez_rubino07}, to a sub-field of the large
field-of-view presented by \cite{orozco_hinode07}
and \cite{lites08} in order to analyze the effect of noise and degeneracies on the retrieved 
parameters.

\section{Observations}
The observations are those analyzed by \cite{lites08}, which were obtained at disk center 
on 2007 March 10 with the spectropolarimeter
aboard \emph{Hinode}. The analyzed map is a small sub-field of 7.4"$\times$7.4" taken from
a large field-of-view of 302"$\times$162". This small sub-field has been also 
analyzed by \cite{orozco_hinode07} to investigate the properties of very quiet internetwork.
The observed spectral region contains the Fe \textsc{i} doublet at 630.1 and 630.2 nm with
a resolution close to 3$\times$10$^5$. The intensity image in the local continuum normalized to the 
average continuum intensity in the sub-field is shown on the left panel
of Fig. \ref{fig:maps} while the total polarization, defined as $\int(Q^2+U^2+V^2)^{1/2} d\lambda/ \langle I \rangle$,
is shown in the right panel of the same figure. This figure also presents the regions of
$\int(Q^2+U^2)^{1/2} d\lambda/ \langle I \rangle > 0.0035$ in red contours and 
$\int(V^2)^{1/2} d\lambda/ \langle I \rangle > 0.003$ in blue contours.
After calibration, the noise level amounts to
1.2$\times$10$^{-3} I_c$ for Stokes $V$ and to 1.1$\times$10$^{-3} I_c$ for Stokes $Q$ and $U$. For more
details on the observations, we refer to \cite{lites08}.

\section{Bayesian Inversion}
\subsection{\B}
Although the \B\ code has been extensively described by \cite{asensio_martinez_rubino07}, we 
give here a brief overview because of its novel characteristics. The \B\ code is an
inversion code built under the framework of the Bayesian approach to inference \citep[see e.g.,][]{neal93,gregory05}. 
Let $M$ be a model that is proposed to explain an observed dataset $D$ and let
$I$ be a set of sensible \emph{a-priori} information about the problem (for instance,
the spectral resolution of the observations). The model
$M$ is parameterized in terms of a vector of parameters of length $N_\mathrm{par}$, $\thetabold  \in \mathbb{R}^{N_\mathrm{par}}$. 
Due to the presence of noise in the observations, any inversion 
procedure is not complete by just giving the values of the model parameters that better 
fit the observations. The full solution to the inference problem is to provide the posterior probability distribution 
function (pdf) $p(\thetabold|D,I)$ that describes the probability that a given set of parameters
is compatible with the observables given the a-priori knowledge. As a consequence, 
statistically relevant information about one parameter (irrespective
of the value taken by other parameters) can be obtained from this pdf by marginalization 
(integration) of the rest of parameters. In order to calculate the posterior 
distribution $p(\thetabold|D,I)$, the Bayes theorem can be applied to obtain:
\begin{equation}
p(\thetabold|D,I) = \frac{p(D|\thetabold,I) p(\thetabold|I)}{p(D|I)},
\label{eq:bayes_theorem}
\end{equation}
where $p(D|I)$ is the so-called \emph{evidence}, a quantity that in the context of parameter
estimation is of no importance (since it is a constant that does not depend on the
model parameters $\thetabold$) but that turns out to be crucial in the context of 
model selection. The distribution $p(\thetabold|I)$ is denominated the prior distribution
and contains all relevant a-priori information about the parameters of the model. Usually,
unless some information is available about the value of some parameters, it is common
to use uninformative (or vague) priors like bounded uniform distributions or Jeffreys' priors
\citep[e.g.,][]{neal93}. Finally, $p(D|\thetabold,I)$ is the likelihood, a distribution
that characterizes how well a model with parameters $\thetabold$ reproduces the
observed dataset. 
The Bayes theorem states that the probability that a model $M$ becomes plausible once the 
information encoded in the data $D$ is taken into account depends on how plausible the model was 
without data and how well the model fits the data.

Assuming that the observables are represented as a vector $\mathbf{d} \in \mathbb{R}^N$, that the
result of evaluating the model $M$ at a given set of parameters $\thetabold$ is the vector
$\mathbf{y}  \in \mathbb{R}^N$ and that the observations are contaminated by a noise component
characterized by the vector $\mathbf{e} \in \mathbb{R}^N$, we have:
\begin{equation}
d_i = y_i + e_i, \qquad \forall i.
\end{equation}
When the chosen model parameters exactly correspond to those of the observed dataset, the 
distribution of differences $y_i-d_i$ has to follow the distribution of the noise. Assuming that 
the noise is Gaussian distributed with a standard deviation given by the vector $\sigmabold \in \mathbb{R}^N$, 
the likelihood function is, therefore, given by the following Gaussian distribution:
\begin{equation}
p(D|\thetabold,I) \propto \prod_{i=1}^N  \exp \left[ \frac{(y_i-d_i)^2}{2\sigma_i^2} \right].
\end{equation}

In the case of \B, the model $M$ is based on the simplified Milne-Eddington (ME) 
description of stellar atmospheres. The emergent Stokes profiles are calculated
as a linear combination of the emergent Stokes profiles obtained from 
an arbitrary number of components, each of them
characterized by the standard ME parameters. Although the Bayesian approach will
allow us in the future to automatically choose the optimal number of components $N_c$
using model selection techniques, $N_c$ is fixed and
chosen a-priori in the present version of \B (in other words, $N_c$ is part of the a-priori information summarized
by $I$).
The vector $\thetabold$ contains all the ME parameters over which we do
inference (the selection of free parameters is completely configurable), while
the remaining parameters ME are set as fixed quantities. The \B\ code neglects the
fixed parameters and they are not included in the vector $\thetabold$, although the result is equivalent to the case
in which the posterior distribution is marginalized (integrated) over them after setting Dirac delta 
priors on them (see \S\ref{sec:marginalization}). The parameters of the ME atmosphere for each component are:
the Doppler width of the line in wavelength units ($\Delta \lambda_\mathrm{dopp}$), 
the line-of-sight component of the macroscopic bulk velocity ($v_\mathrm{LOS}$), the gradient of the source 
function ($\beta$), the ratio between the line and continuum absorption coefficients ($\eta_0$), 
a line damping parameter ($a$) and the magnetic field vector parameterized by its strength, inclination and azimuth
with respect to a given reference direction ($B$, $\theta_B$ and $\chi_B$, respectively). We
refer, somewhat arbitrarily, to $\Delta \lambda_\mathrm{dopp}$, $v_\mathrm{LOS}$, $\beta$, $\eta_0$ and $a$ as
the thermodynamical parameters (although their pure thermodynamical character is quite diffuse). Additionally, 
we take into account the filling factor of each component, $f_i$, subject to the constraint $\sum_{i=1}^{N_c} f_i=1$.
The maximum number of free parameters is $9N_c-1$.

In order to sample the posterior distribution function and carry out the ensuing marginalizations,
\B\ utilizes a Markov Chain Monte Carlo \citep[MCMC;][]{metropolis53,neal93} scheme based on the
Metropolis algorithm. The initial proposal density distribution is a multivariate Gaussian
with diagonal covariance matrix that is set to 10\% of the allowed range of variation of
the parameters. After a configurable initial period, the proposal density is changed
to a multivariate Gaussian with a covariance matrix that is estimated from the previous
steps of the chain\footnote{This is an improvement over the version of \B\ presented by \cite{asensio_martinez_rubino07}
which used a proposal density with diagonal covariance matrix.}. In order to improve convergence, the covariance matrix is multiplied
by a quantity that assures that the acceptance rate of proposed models is close to 25\%, a 
value that is the theoretical optimal value for simple problems \citep{gelman96}.

\begin{figure}
\plotone{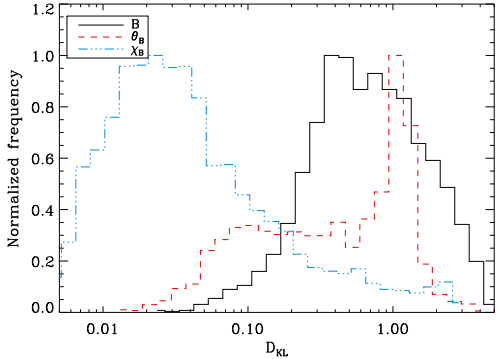}
\caption{Histogram of Kullback-Leibler divergence between the posterior and prior distributions
for the magnetic field strength (black line), inclination (red line) and azimuth (blue line). 
This indicates that the amount of information present in the observations is sufficient
to return posterior distributions clearly different from the prior distributions for the
magnetic field strength and for the inclination. The information is heavily reduced for the 
azimuth of the field.}
\label{fig:kl_histogram}
\end{figure}

\subsection{Marginalization}
\label{sec:marginalization}
In order to obtain the posterior probability distribution function for
one parameter and give estimations and confidence intervals, we have
to marginalize (integrate out) the rest of parameters:
\begin{equation}
p(\theta_i|D,I) = \int \mathrm{d}\theta_1\mathrm{d}\theta_2 \cdots \mathrm{d}\theta_{i-1} \mathrm{d}\theta_{i+1} \cdots 
\mathrm{d}\theta_{N_\mathrm{par}} p(\thetabold|D,I).
\end{equation}
Interestingly, the output of any MCMC code is a converged Markov chain for each parameter 
whose histogram is just proportional to $p(\theta_i|D,I)$. Therefore, calculating marginal
posterior distributions is just a matter of making histograms. Any relevant statistical
information can be obtained from the distribution or summarized using, in the appropriate
cases, the median and confidence intervals or upper/lower limits.

\subsection{This analysis}
Due to the general character of \B, it is possible to adapt it to special inversion
schemes like that applied by \cite{orozco_hinode07} for the inversion of \emph{Hinode}
quiet-Sun internetwork magnetic fields. We slightly modified \B\ to suit the
scheme followed by these authors since our aim is to analyze the reliability of 
their results. The model used consists of one magnetic Milne-Eddington component
contaminated by the presence of a stray-light component. This stray-light
contamination, characterized by a filling factor $\alpha$ is obtained locally as the
average of the pixels within a box of 1" centered in the pixel of interest. The 
magnetic filling factor (fraction of signal represented by a magnetic atmosphere) is,
therefore, $1-\alpha$. In the analysis, we focus on the magnetic filling factor.

A direct consequence of using a Bayesian inference approach is that we do not need to
establish a selection criterion to choose which pixels to invert. For those pixels
where the signal-to-noise level is too low, the ensuing posterior probability distributions will
straightforwardly show the degeneracies present in the data and the resulting
posterior distributions will automatically be very similar to the prior distributions. Therefore, we invert all
the pixels in the field-of-view and select afterwards which pixels provide a sufficient
amount of information to the inference process. Apart from the pure visual inspection of
the resulting marginal probability distributions, we use a criterium to characterize 
the pixels in which there is sufficient information. The criterium is based on the
calculation of the Kullback-Leibler divergence \citep[KLD;][]{kullback_leibler51} between
the two probability distributions $p(\theta|D,I)$ (posterior distribution) and
$p(\theta|I)$ (prior distribution):
\begin{equation}
D_\mathrm{KL} = \int_{-\infty}^\infty p(\theta|D,I) \log_2 \frac{p(\theta|D,I)}{p(\theta|I)} \mathrm{d}\theta.
\end{equation}
The KLD, always a positive quantity, measures the difference in number of bits 
between transmitting samples of the distribution $p(\theta|D,I)$ 
using a code based on $p(\theta|D,I)$ and transmitting samples of the distribution $p(\theta|D,I)$
using a code based on the distribution $p(\theta|I)$. The more different
both distributions are, the larger the additional information required. Therefore, larger 
values of $D_\mathrm{KL}$ indicate that both distributions are very different. We apply this
measure to compare the final posterior probability distribution after the data has been presented
to the inference code and the prior distribution. When the data is informative, the posterior
distribution is very different from the prior distribution and $D_\mathrm{KL}$ will be large.
Since the priors for the stray-light factor, magnetic field strength, inclination 
and azimuth of the field are uniform distributions in the intervals $[0,1]$, $[0,1500]$ G, $[0^\circ,180^\circ]$ and
$[0^\circ,180^\circ]$, respectively, the KLD simplifies to:
\begin{equation}
D_\mathrm{KL} = \log_2 (\theta_\mathrm{max}-\theta_\mathrm{min}) + \int_{-\infty}^\infty p(\theta|D,I) \log_2 p(\theta|D,I) \mathrm{d}\theta,
\end{equation}
where the last term represents the negative of the standard definition of the entropy (sometimes called negentropy)
of a continuous probability distribution function.
This shows that the limit $D_\mathrm{KL}=0$ is reached when the posterior distribution is 
equal to the prior distribution and that it can potentially reach very high values for
very informative posterior distributions (because the entropy is not bounded from above). 
According to the results presented below, pixels with $D_\mathrm{KL} \gtrsim 0.5$ can
be considered to have enough information to roughly constrain a given parameter, although 
a more conservative threshold would be $D_\mathrm{KL} \gtrsim 1$.


\section{Results and discussion}
After applying the \B\ code to all the pixels in the field-of-view, the results consist of converged Markov
chains for each parameter. As an additional test for convergence, we rerun the code
several times obtaining almost indistinguishable results.
By making histograms of each chain, we obtain the posterior distribution function
marginalized over all parameters except the one of interest. These marginal posterior distribution
functions show how the information encoded on the observed Stokes parameters constrain
every individual parameter. In principle, the correct answer to the
inversion process would be to give such posterior distributions for each parameter and each pixel.
Since this is obviously not feasible for presentation purposes, we follow a different
approach and try to condense the important statistical information present in each chain. 
We focus only on the properties of the magnetic field vector thanks to the results of \cite{westendorp98},
who demonstrated that the cross-talk between thermodynamical and magnetic parameters is of reduced importance
in Milne-Eddington inversions. The reason for this behavior has to be found on the large
difference between the response functions for the Milne-Eddington thermodynamical and magnetic
parameters, as elegantly shown by \cite{orozco_deltoro07}.
Note that, due to the Bayesian approach that we follow, the marginal posterior distributions for the 
magnetic field parameters automatically include information about all possible values of these 
thermodynamical parameters weighted by their probability.

\begin{figure*}
\includegraphics[width=0.5\textwidth]{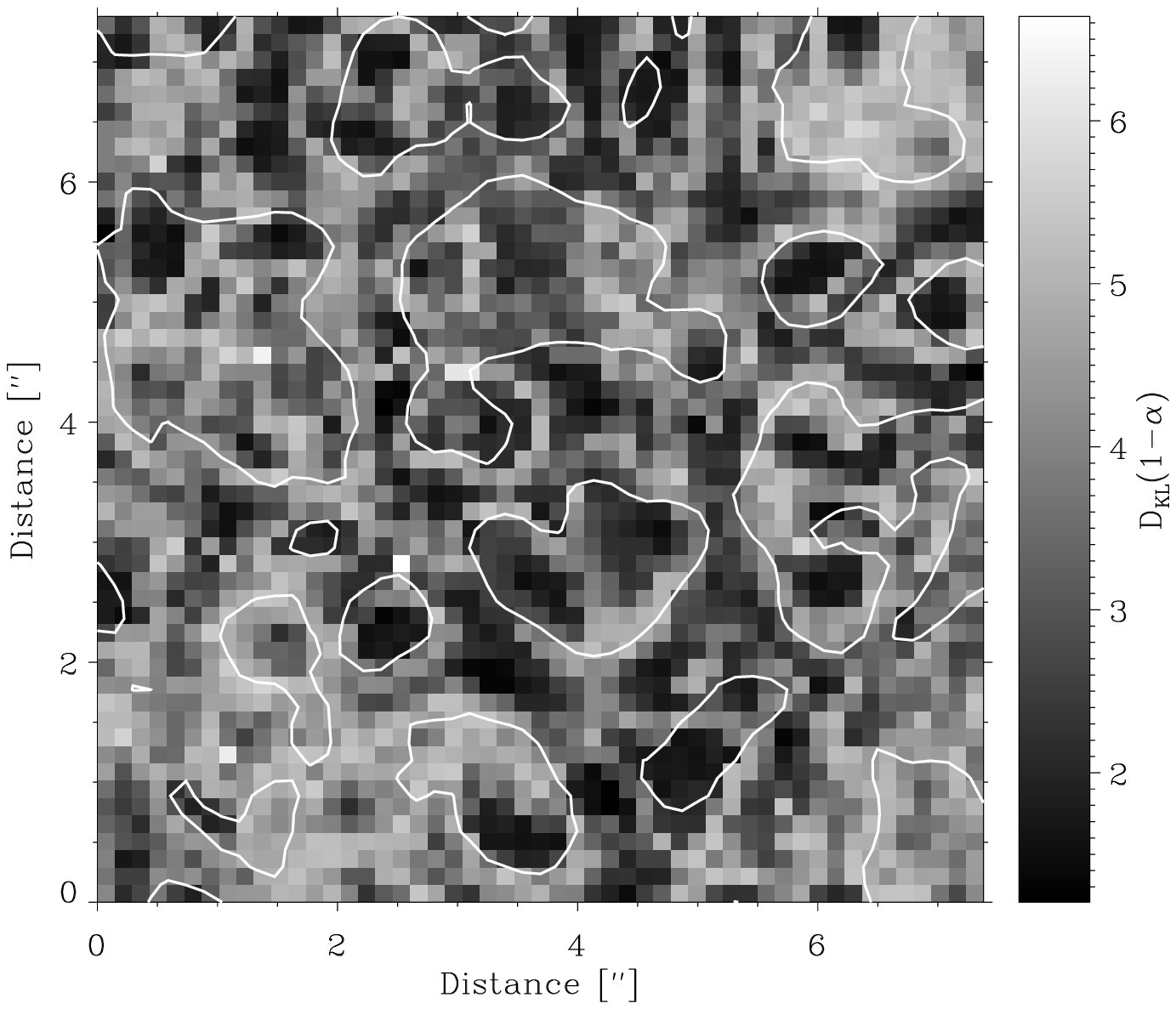}%
\includegraphics[width=0.5\textwidth]{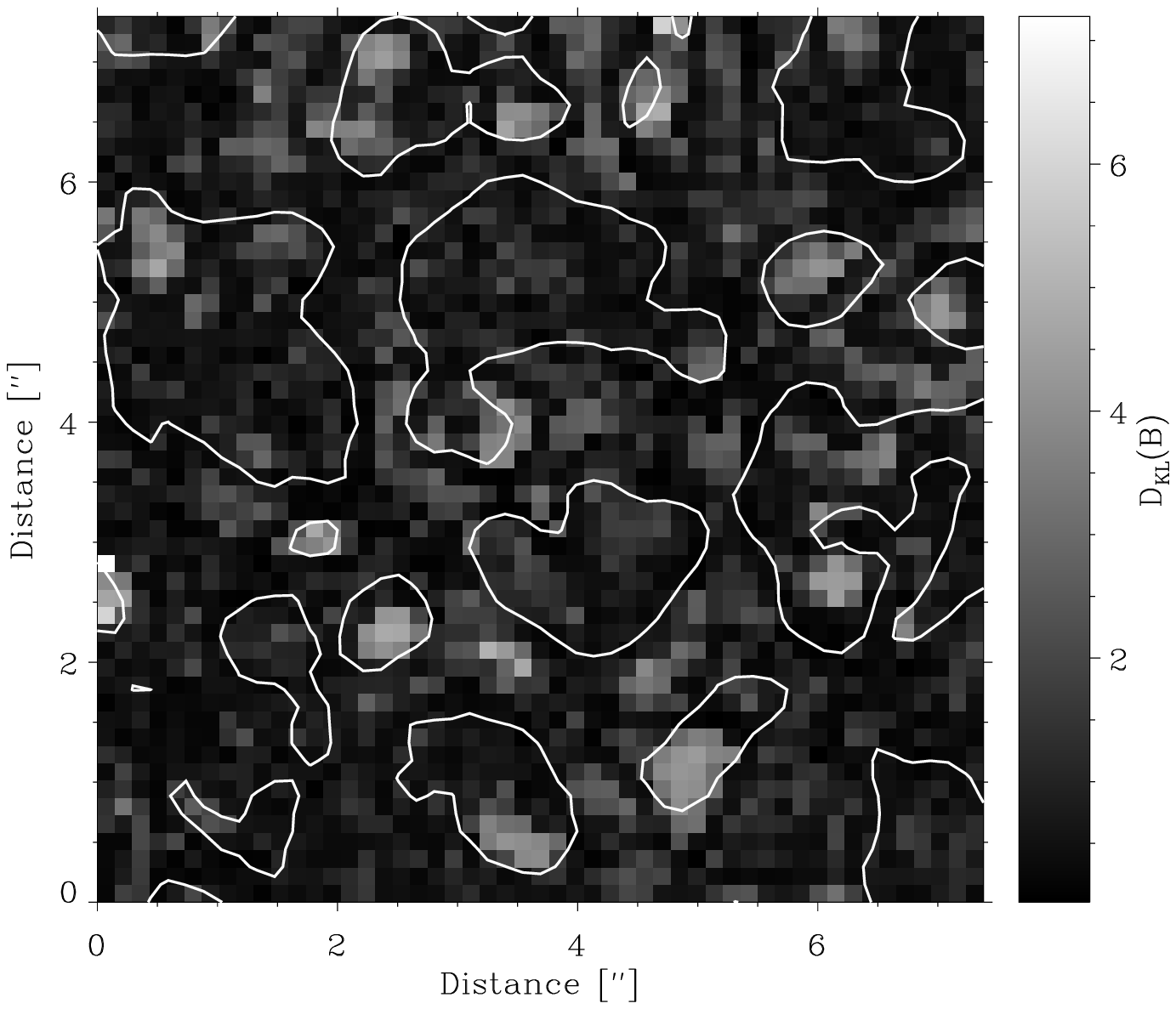}
\includegraphics[width=0.5\textwidth]{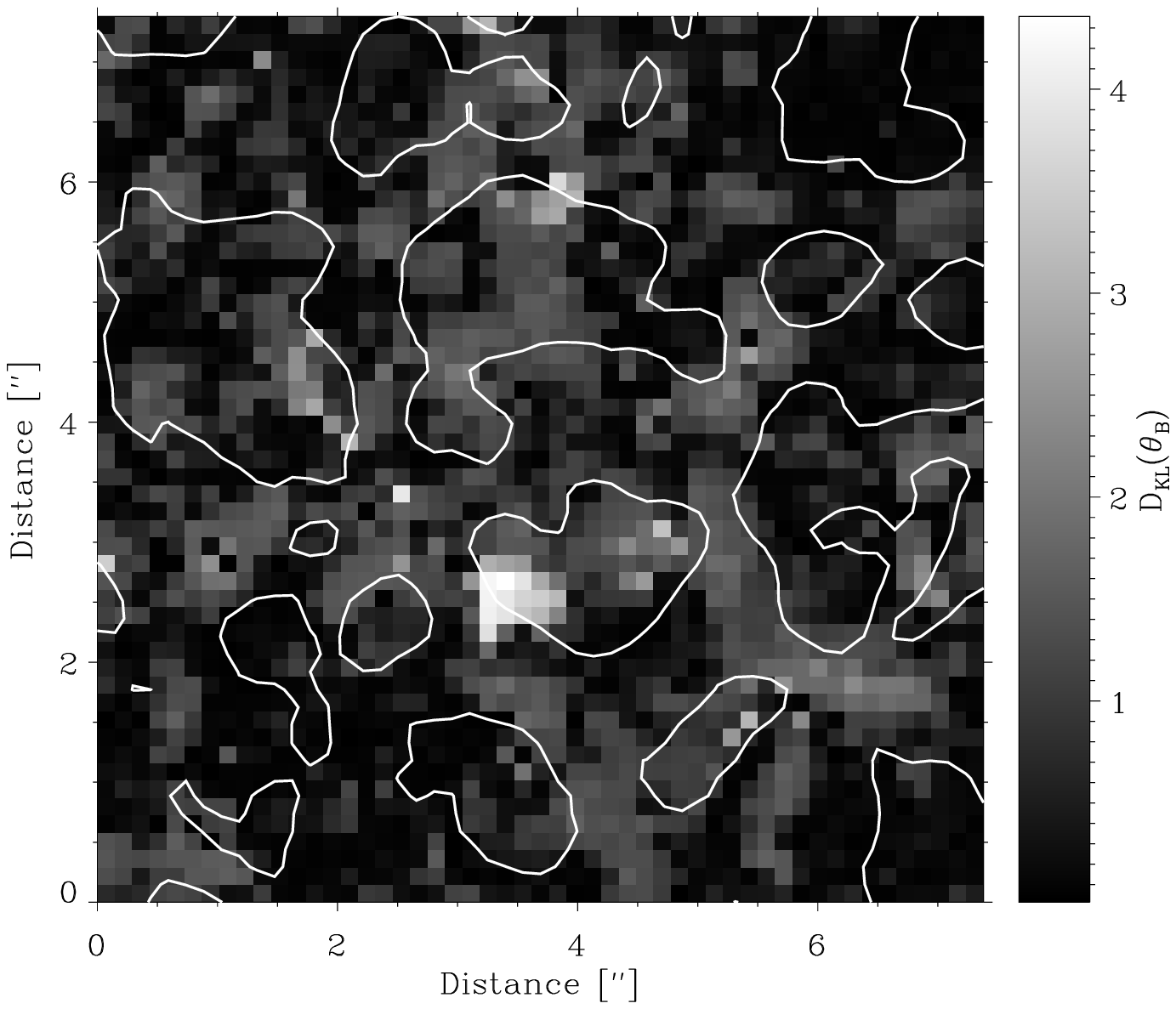}%
\includegraphics[width=0.5\textwidth]{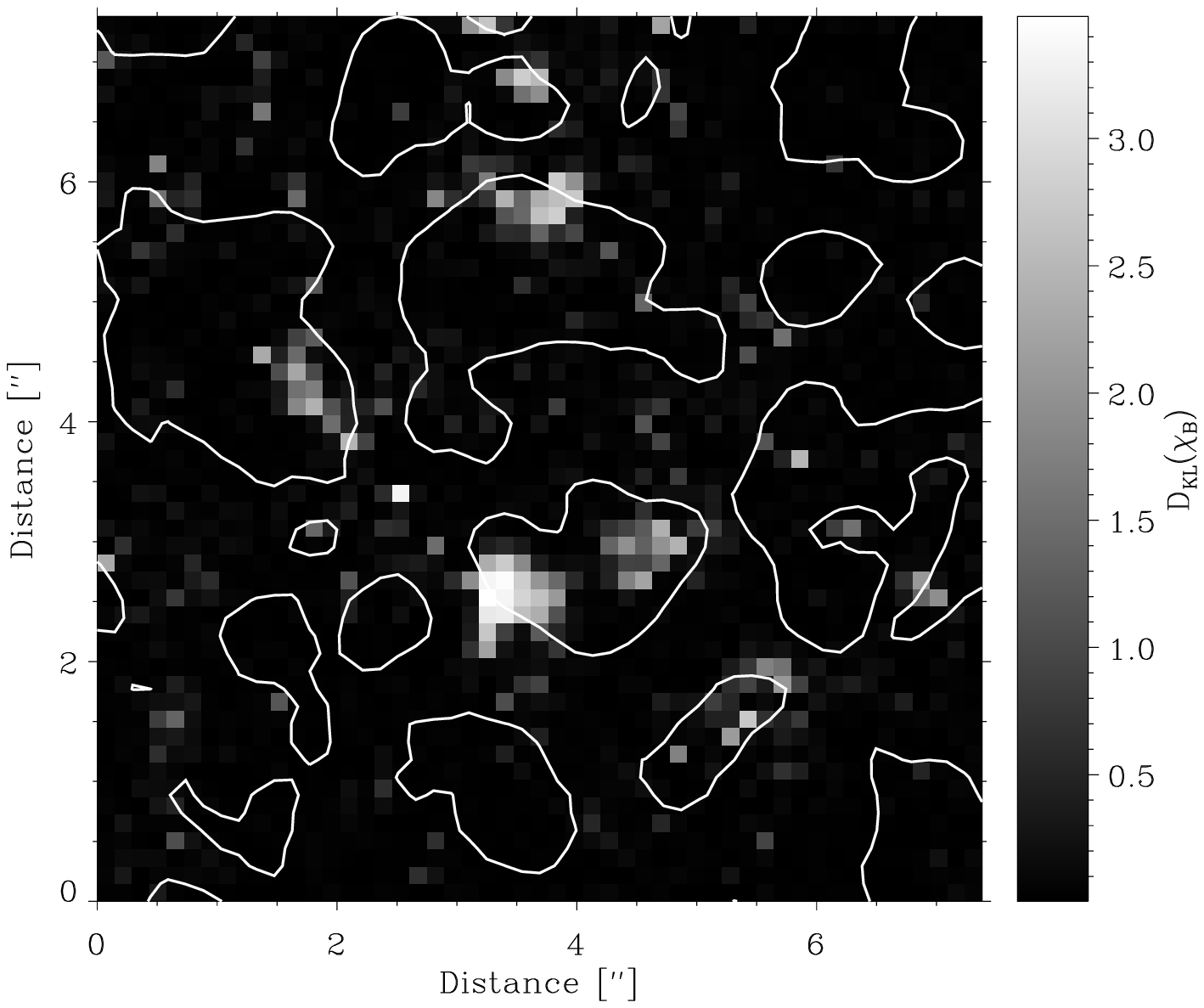}
\includegraphics[width=0.5\textwidth]{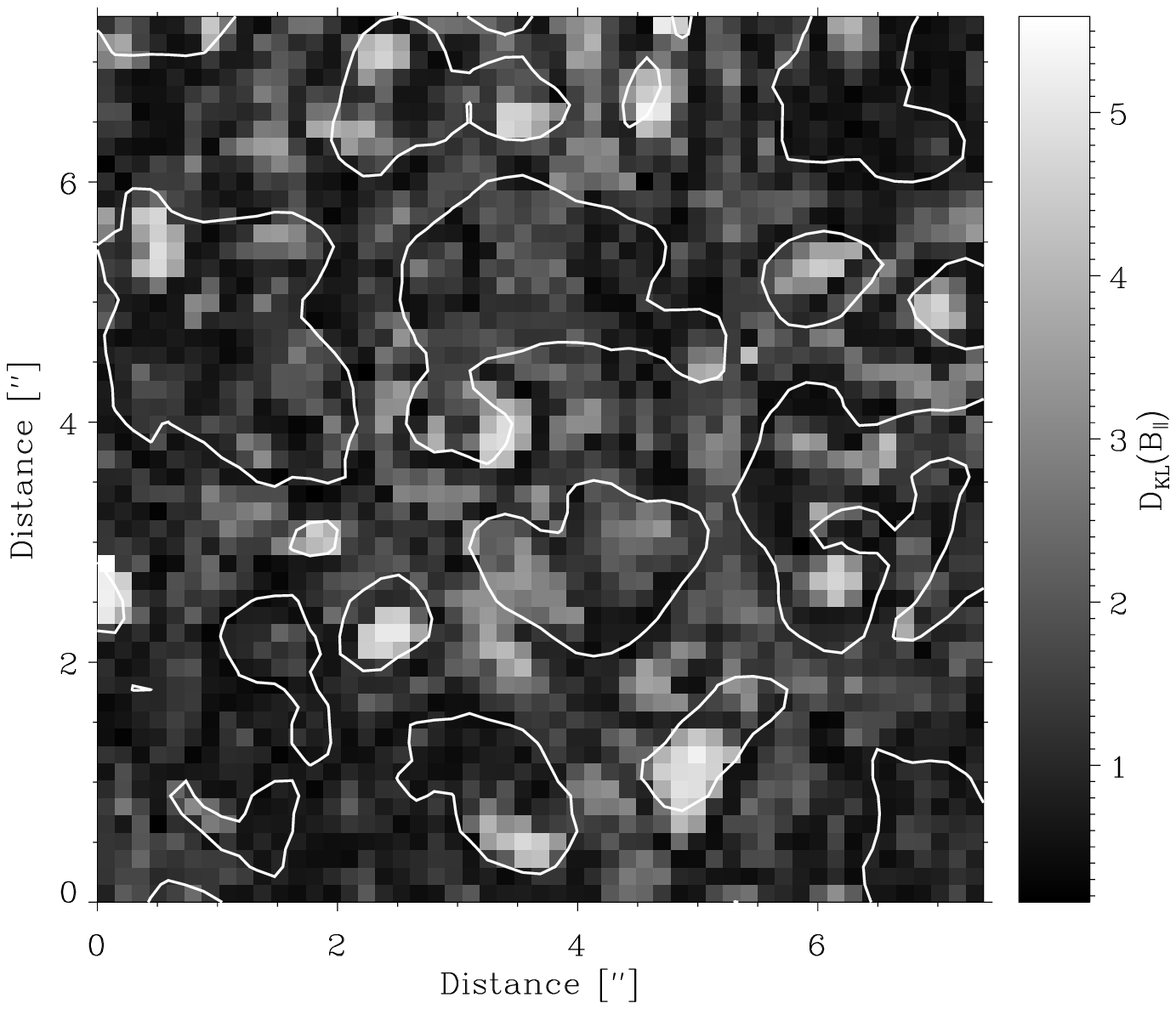}%
\includegraphics[width=0.5\textwidth]{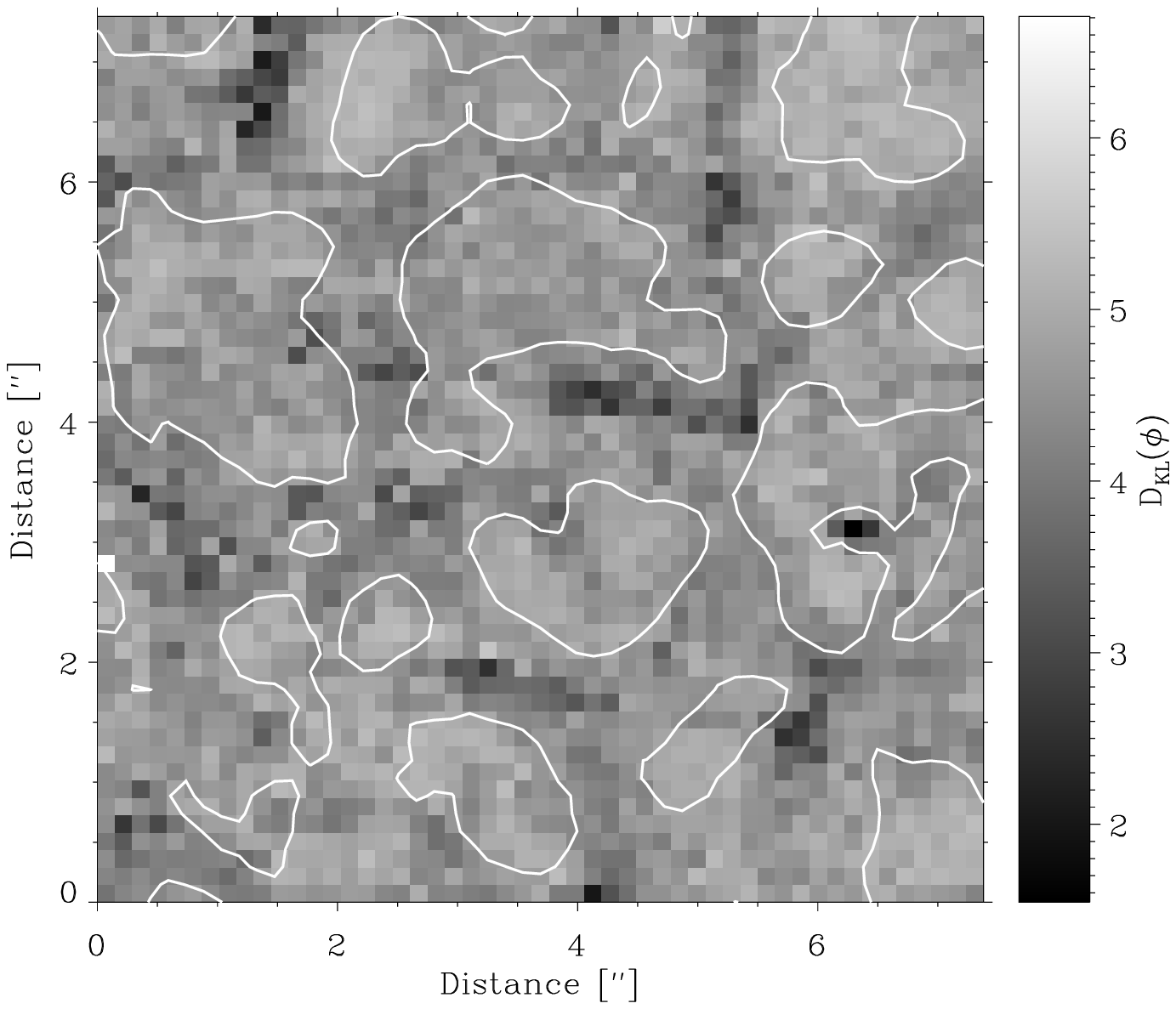}
\caption{Horizontal variation of the Kullback-Leibler divergence between the posterior
and prior distributions for the magnetic filling factor (upper left panel), the magnetic
field strength (upper right panel), field inclination (middle left panel), 
field azimuth (middle right panel), longitudinal component of the field (lower left panel)
and longitudinal magnetic flux density (lower right panel). The contours indicate regions where $I_c/\langle I_c \rangle=1.02$.}
\label{fig:kl_images}
\end{figure*}

\subsection{Amount of information}
The first key point is to estimate how much information about every parameter can be inferred from the
observables. To this end, we analyze the Kullback-Leibler divergences. Figure \ref{fig:kl_histogram}
shows the histogram of $D_\mathrm{KL}$ for the marginal posterior distributions of the
magnetic field strength, inclination and azimuth in all pixels. It is clear from this figure that
the amount of information available to constrain the azimuth is very poor because $D_\mathrm{KL}$
is very close to zero for all the pixels. This is a consequence of the fact that almost all pixels
present posterior distributions for the azimuth that are indistinguishable from the uniform prior 
distribution, as we demonstrate below. There is more information available for constraining the field strength
and inclination (larger values of $D_\mathrm{KL}$) although several pixels present posterior 
distributions that are still similar to the prior. Table \ref{tab:kl_divergences}
summarizes the percentage of pixels with relevant information for each parameter in the field-of-view.

Of interest is also the analysis of the horizontal variation of $D_\mathrm{KL}$ 
shown in Fig. \ref{fig:kl_images}. The results clearly indicate what has been
summarized in Fig. \ref{fig:kl_histogram}. For clarity, we have also indicated
with contours the regions where $I_c/\langle I_c \rangle=1.02$, which may
serve as a rough indication of the position of the brighter parts of
granules. Several interesting points can be inferred from these plots.

First, a large amount of pixels present large values of $D_\mathrm{KL}(1-\alpha)$ 
(bright points in the upper left panel of the figure). These pixels
have, as we show later in Fig. \ref{fig:posterior_examples_kl}, 
posterior distributions for the magnetic filling factor with a very 
conspicuous peak close to zero. They correspond to pixels in which
the polarization signal is extremely low and well below the noise. In such
a situation, and since the intensity profile is non-zero, the best
fit is accomplished with $\sim$100\%
stray-light contamination. For this reason, the very same white pixels for
the $D_\mathrm{KL}(1-\alpha)$ map are typically dark in the $D_\mathrm{KL}(B)$ map
because the noise rapidly destroys the information 
needed to give a reliable value of the magnetic field strength. 
It is important to point out that these results are only valid for the present noise
level of the observations and that results would surely change if a better
signal-to-noise ratio had been achieved. 


Second, it is obvious from the middle panels of Fig. \ref{fig:kl_images} that
there is less information to constrain the inclination and azimuth of the field 
(especially for the azimuth). However, it is interesting to point out that there is apparently
more information available in intergranular lanes than in granules, in many cases 
corresponding to pixels with the largest polarization signals. These polarization
signals in the integranular lanes are mainly circular polarization, as shown
in Fig. \ref{fig:maps}. It is also obvious from
Fig. \ref{fig:kl_images} the appearance of patches in the field-of-view where the field strength,
inclination and azimuth can be nicely constrained. Note that, according to Fig. \ref{fig:maps},
they correspond to patches where the linear polarization signal is large. This demonstrates,
as we shown in \S\ref{sec:isotropy} and Fig. \ref{fig:posterior_theta_signalQU},
the well-known fact that a linear polarization signal well above the noise level produces a 
well constrained inferred magnetic field vector. Particularly frustrating is
the fact that, at this noise level and apart from these patches, the information for constraining 
the azimuth of the field is almost negligible.

Finally, the lower panels present the horizontal variation of the Kullback-Leibler divergence 
for $B_\parallel$, the longitudinal component of the magnetic field vector (lower left panel)
and $\phi=(1-\alpha)B_\parallel$, the longitudinal magnetic flux density (lower right panel). The information available
for the line-of-sight component of the magnetic field vector is very similar to that of
the field itself but slightly larger because it is augmented with the knowledge of the
inclination angle. On the contrary, the longitudinal magnetic flux density shows a very clear
correlation with the granulation pattern. All values of $D_\mathrm{KL}(\phi)$
are clearly larger than 1, showing that there is sufficient information in all the
pixels to obtain information about this quantity, which clearly changes smoothly
from pixel to pixel.

\begin{table}
\caption{Pixels with Kullback-Leibler divergences above different thresholds.}
\label{tab:kl_divergences}
\centering
\begin{tabular}{c c c c}
\hline\hline
Parameter & $N(D_\mathrm{KL}) > 0.5$ & $N(D_\mathrm{KL}) > 1$  & $N(D_\mathrm{KL}) > 2.5$\\
\hline
$\alpha$ & 100.0\% & 100.0\% & 77.5\% \\
$B$      & 64.8\% & 36.9\% & 9.2 \% \\
$\theta_B$ & 49.8\% & 33.7\% & 1.5\%\\
$\chi_B$  & 7.0\% & 3.8\% & 0.8\% \\
$B_\parallel$ & 90.6\% & 61.0\% & 14.1\% \\
$\phi$ & 100.0\% & 100.0\% & 99.7\%\\
\hline
\end{tabular}
\end{table}

\subsection{Marginal posteriors}
\begin{figure*}
\includegraphics[width=0.5\textwidth]{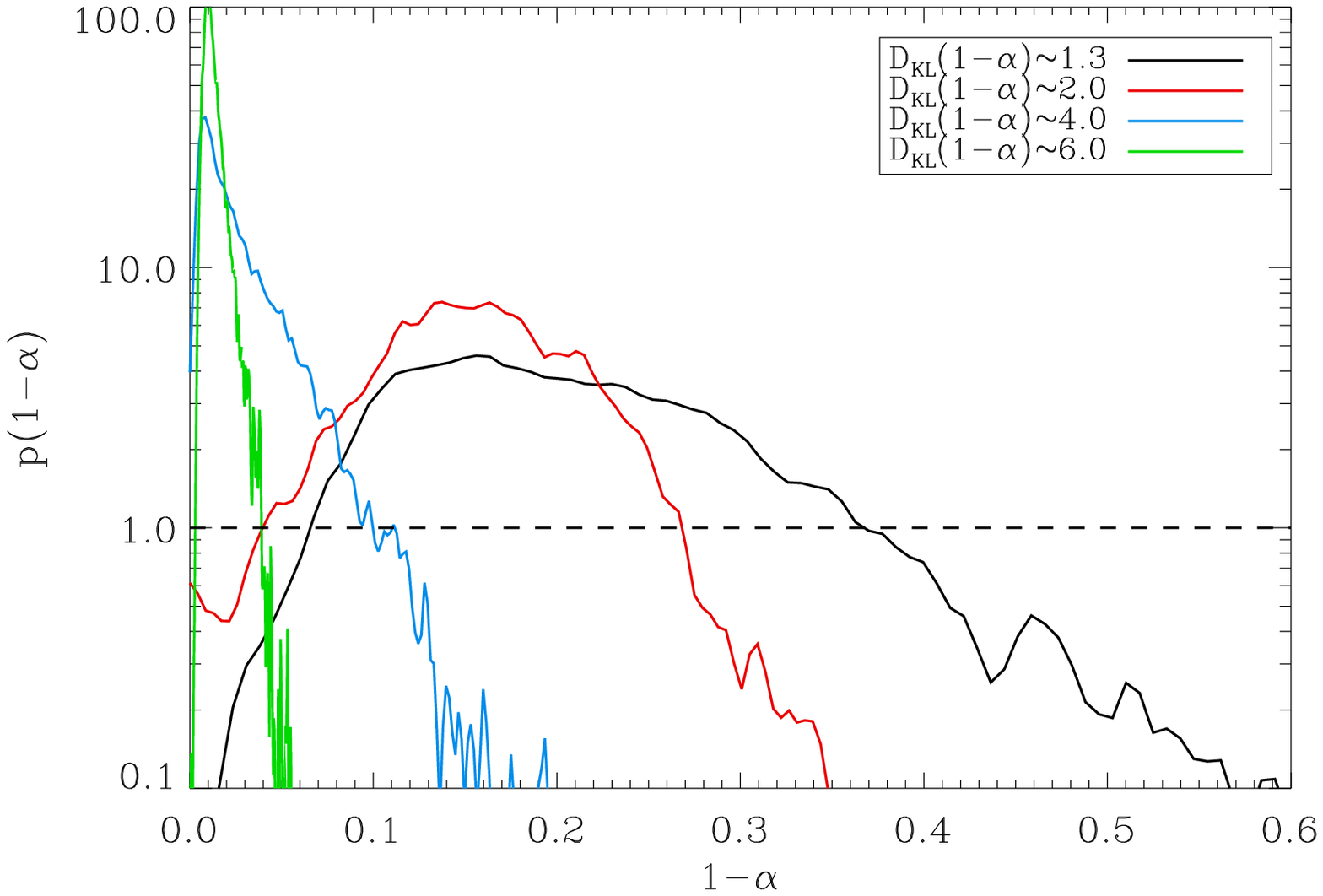}%
\includegraphics[width=0.5\textwidth]{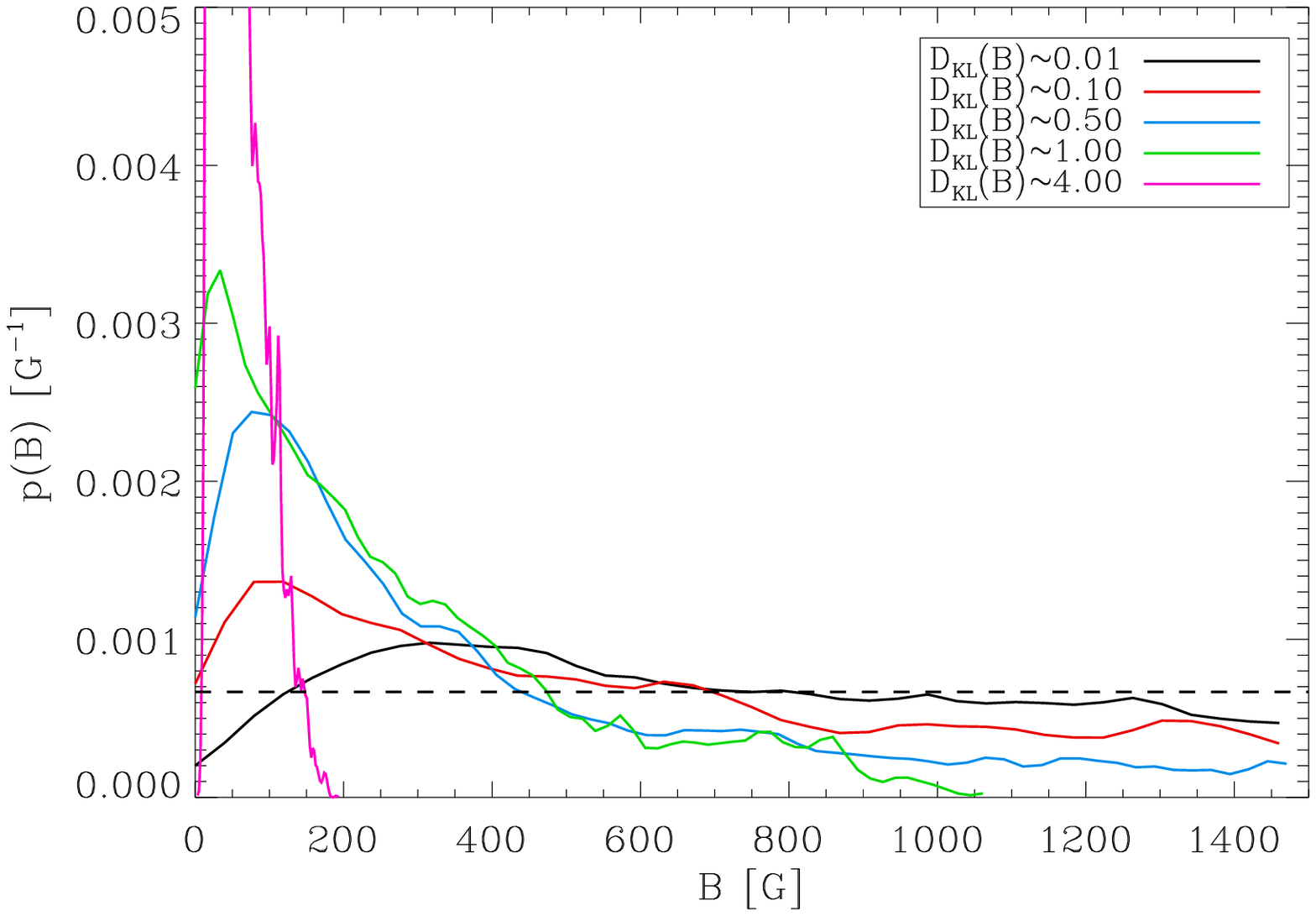}
\includegraphics[width=0.5\textwidth]{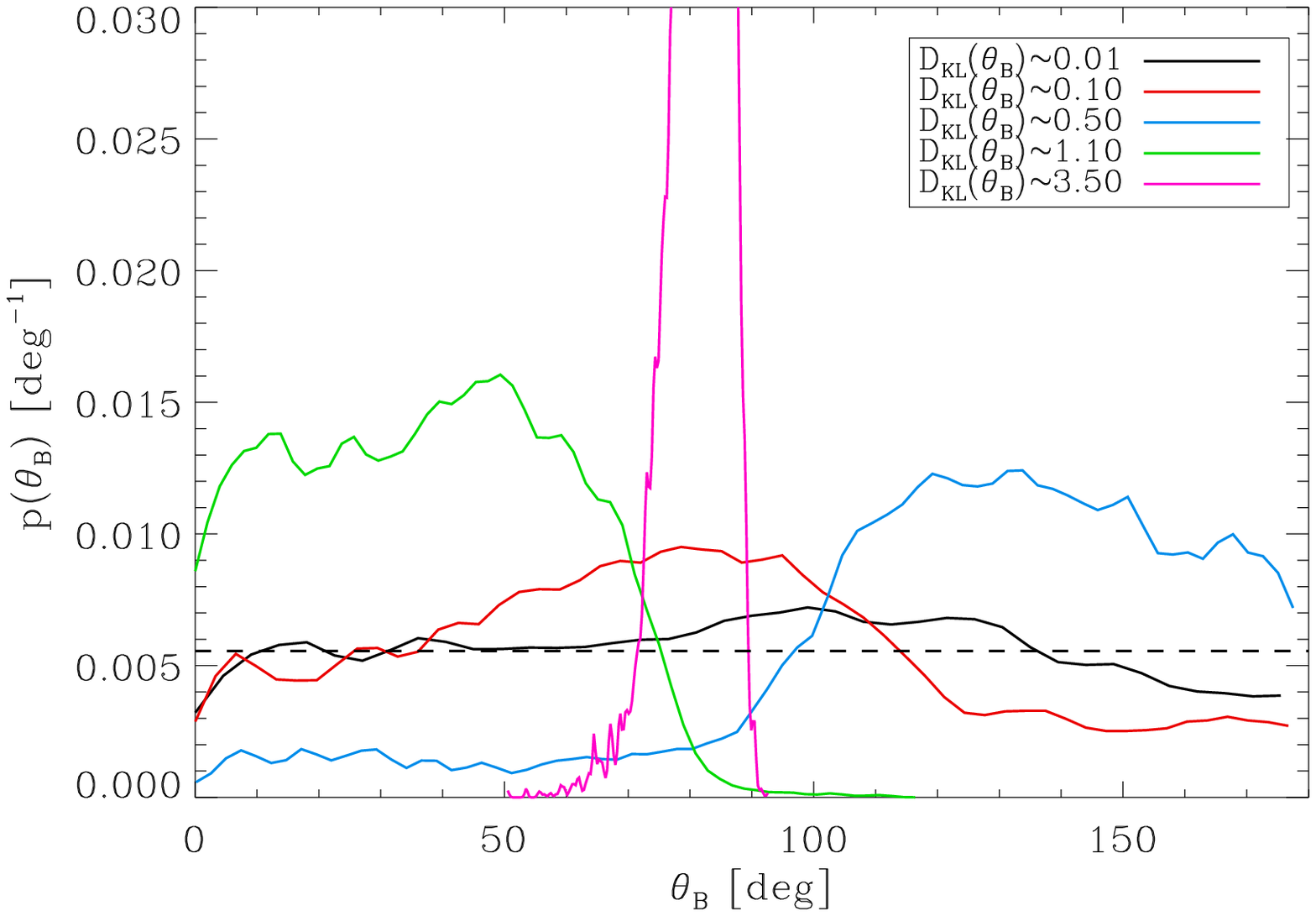}%
\includegraphics[width=0.5\textwidth]{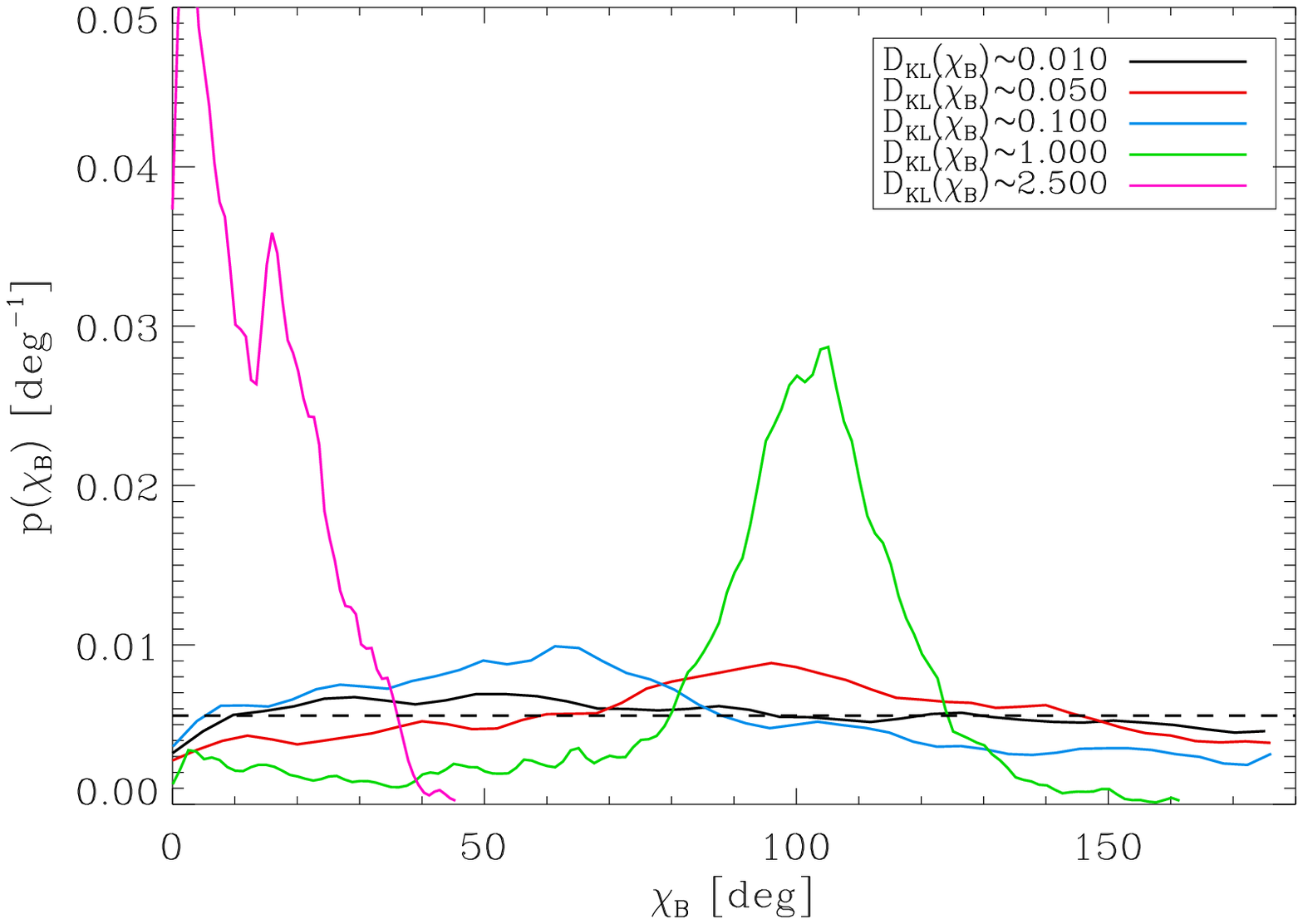}
\includegraphics[width=0.5\textwidth]{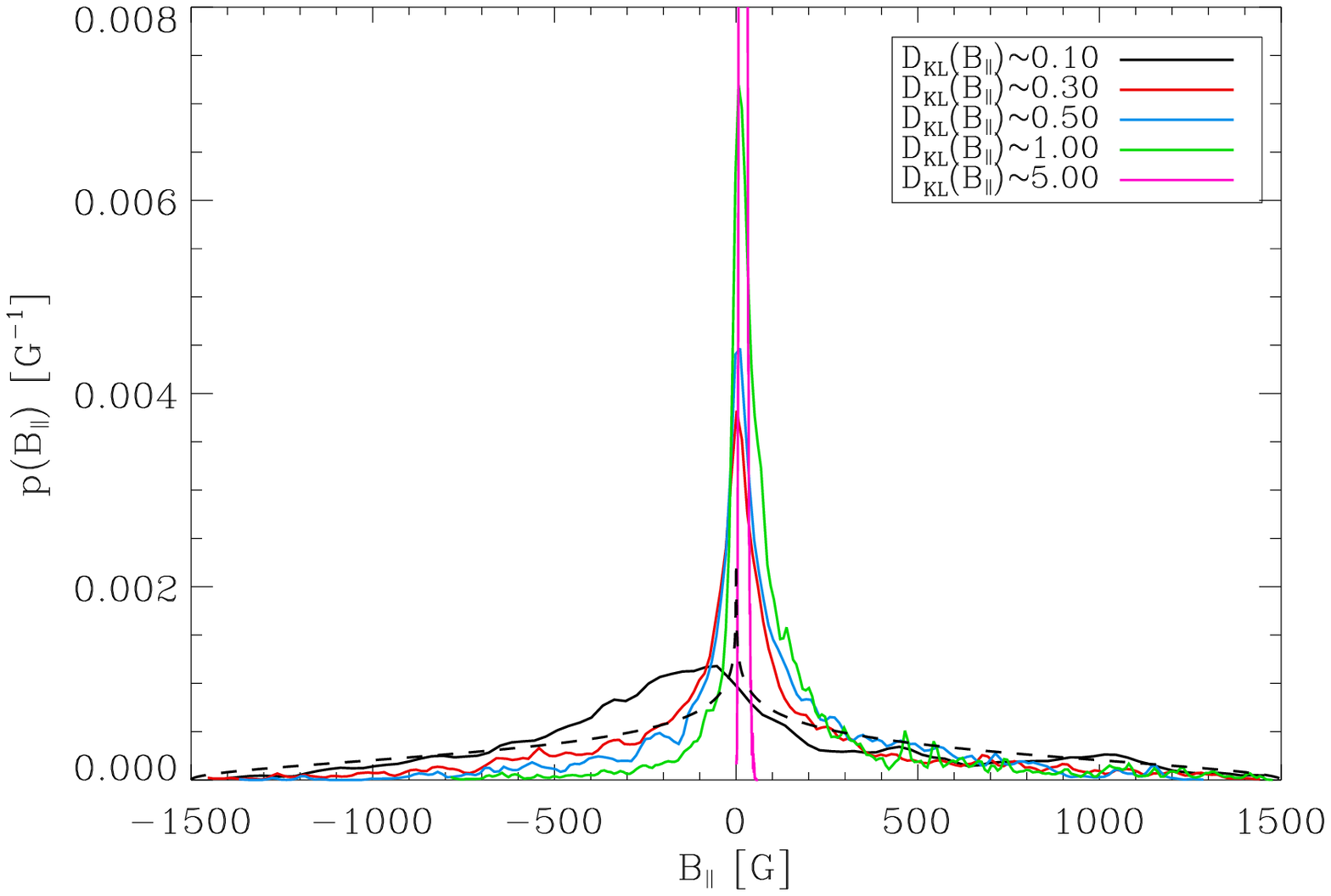}%
\includegraphics[width=0.5\textwidth]{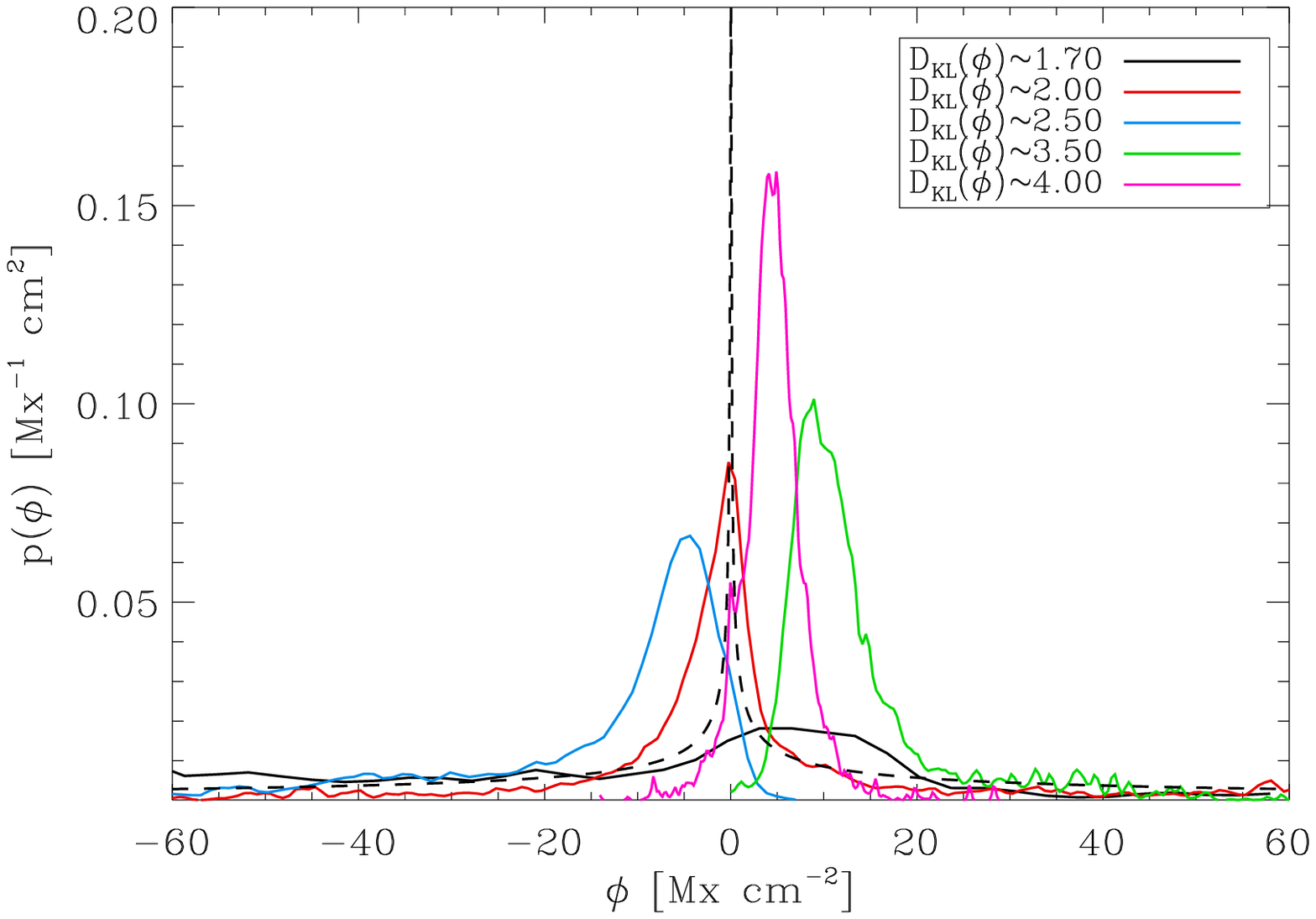}
\caption{Marginal posterior distribution functions for the magnetic filling factor (upper left panel), magnetic field
strength (upper right panel), field inclination (middle left panel), field azimuth (middle right
panel), longitudinal component of the field vector (lower left panel) and longitudinal magnetic flux density(lower
right panel) for several pixels with different values of
the Kullback-Leibler divergence. The selected pixels are representative of zones where there is no
information about the parameters (small values of $D_\mathrm{KL}$) and those in which the 
information is enough to restrict to some extent their values (large values of $D_\mathrm{KL}$). The posterior
distributions have been smoothed with a Gaussian kernel for aesthetic purposes. For comparison, the 
prior distribution is shown in dashed lines.}
\label{fig:posterior_examples_kl}
\end{figure*}

In order to understand the typical values of $D_\mathrm{KL}$ for which we can
consider that the posterior distribution is clearly different from the prior
distribution, we show in Fig. \ref{fig:posterior_examples_kl} examples of the 
posterior distributions for the magnetic filling factor, magnetic field strength, 
inclination and azimuth for different values of $D_\mathrm{KL}$. We also present
the posterior distributions for the longitudinal component of the field
and for the longitudinal magnetic flux density. They show, in
general, that nicely constrained parameters present $D_\mathrm{KL} \ga 1$. In order
to facilitate the comparison, we also overplot the prior distribution in dashed lines.

As already discussed, the stray-light contamination is usually well constrained, at least
for values of $D_\mathrm{KL}(1-\alpha) \geq 1.3$ (see upper left panel of
Fig. \ref{fig:posterior_examples_kl}). Smaller values present posterior
distributions that can be considered commensurate with the prior distribution. Intermediate
values typically correspond to those points in which the stray-light contamination
amounts to $\alpha \sim 0.8$ on average, while the Bayesian results for those 
points with large values of $D_\mathrm{KL}(1-\alpha)$ point to an 
almost non-magnetic atmosphere. As already discussed by \cite{orozco_pasj07}, this
behavior is a consequence of the fact that the stray-light contamination is mainly sensitive
to the Stokes $I$ profile and can be constrained fairly well irrespective of the noise level
in the rest of Stokes parameters.

A similar behavior is seen in the marginal posteriors for the magnetic field strength (upper
right panel of Fig. \ref{fig:posterior_examples_kl}),
although the results are clearly less informative. When $D_\mathrm{KL}(B) \la 0.1$, we can
state that the posterior distribution is very similar to the prior distribution (marked as
a horizontal dashed line), so
that these pixels do not contain enough information for constraining this parameter. When
$D_\mathrm{KL}(B)$ increases, the posterior distribution functions usually show the same
behavior, peaking at magnetic fields close to zero but with an extended tail towards higher
values. For such distributions with extended tails, the only quantity with statistical meaning that we can 
give to summarize the results is an upper limit (with 68\% or 95\% confidence).
Although an analysis of these upper limits will be presented in \S\ref{sec:parameter_estimation}, it is
possible to see from the upper right panel of Fig. \ref{fig:posterior_examples_kl} that 
they are systematically favouring hG fields with respect to stronger fields.
The set of points associated to the bright patch of the middle panels of Fig. \ref{fig:kl_images}
produce marginal posterior distributions with a Gaussian-like
shape peaking in the 400-600 G range but with tails that completely discard fields in the kG regime. This
represents one of the key conclusions of our work.

\begin{figure*}
\includegraphics[width=0.5\textwidth]{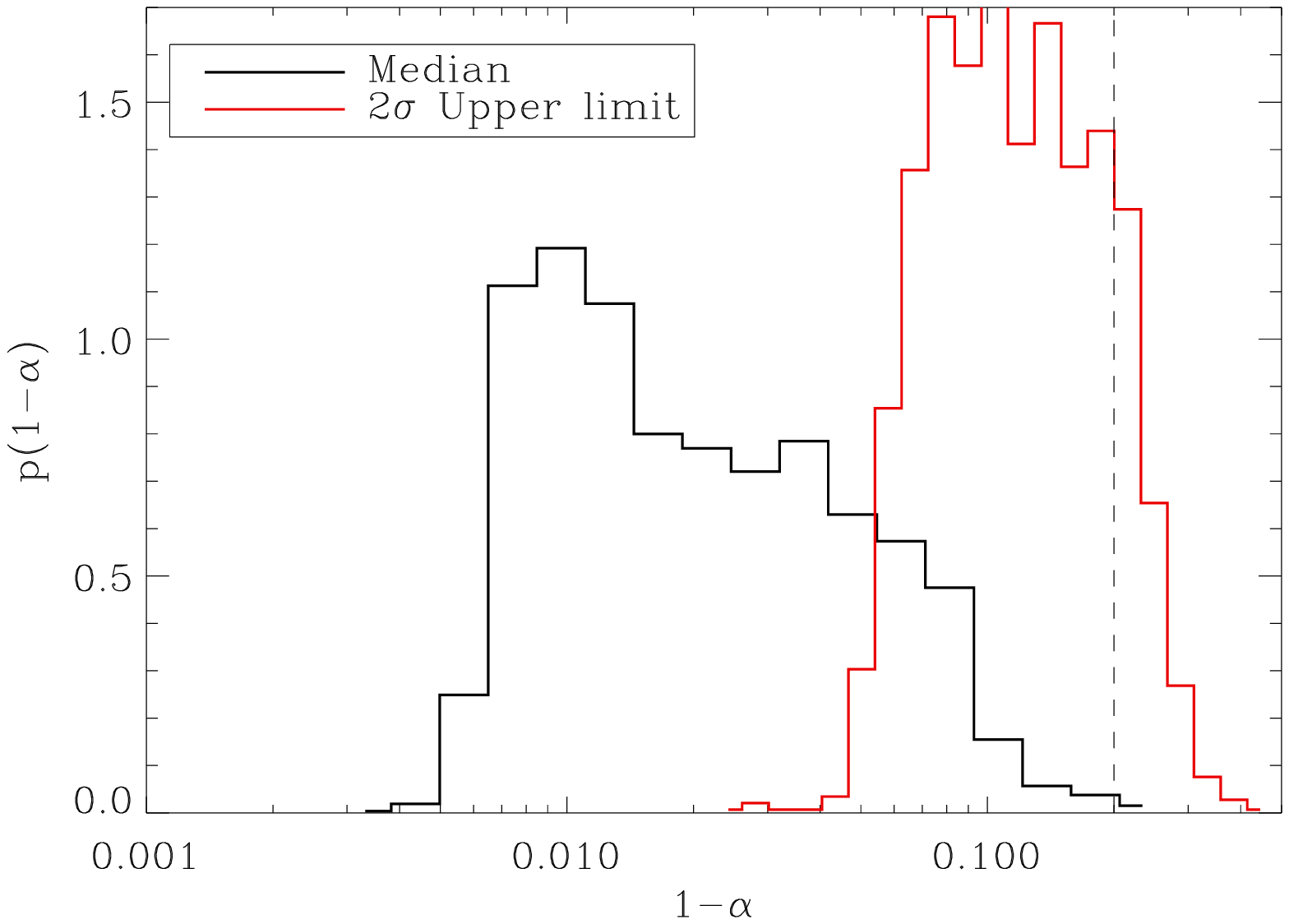}
\includegraphics[width=0.5\textwidth]{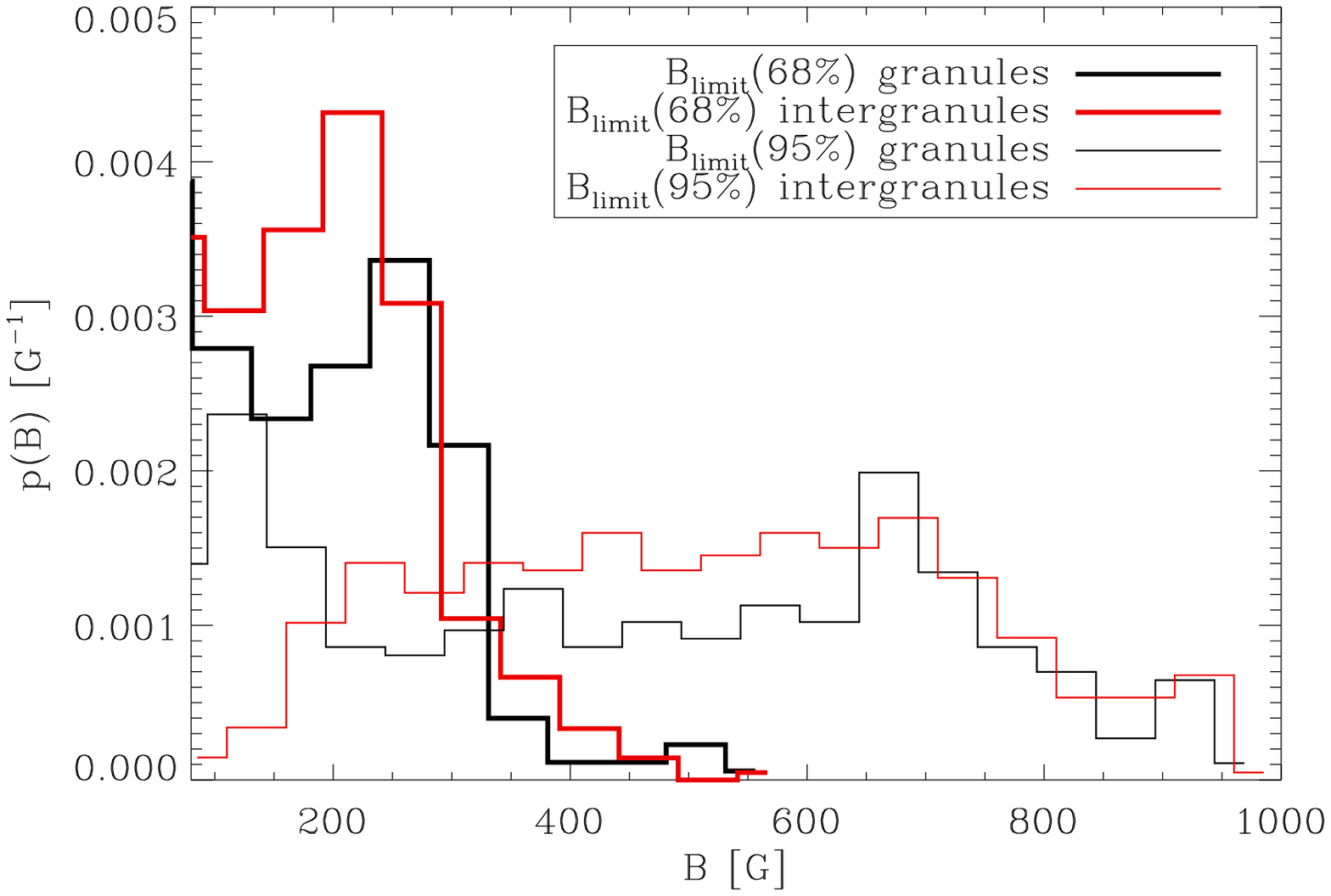}
\includegraphics[width=0.5\textwidth]{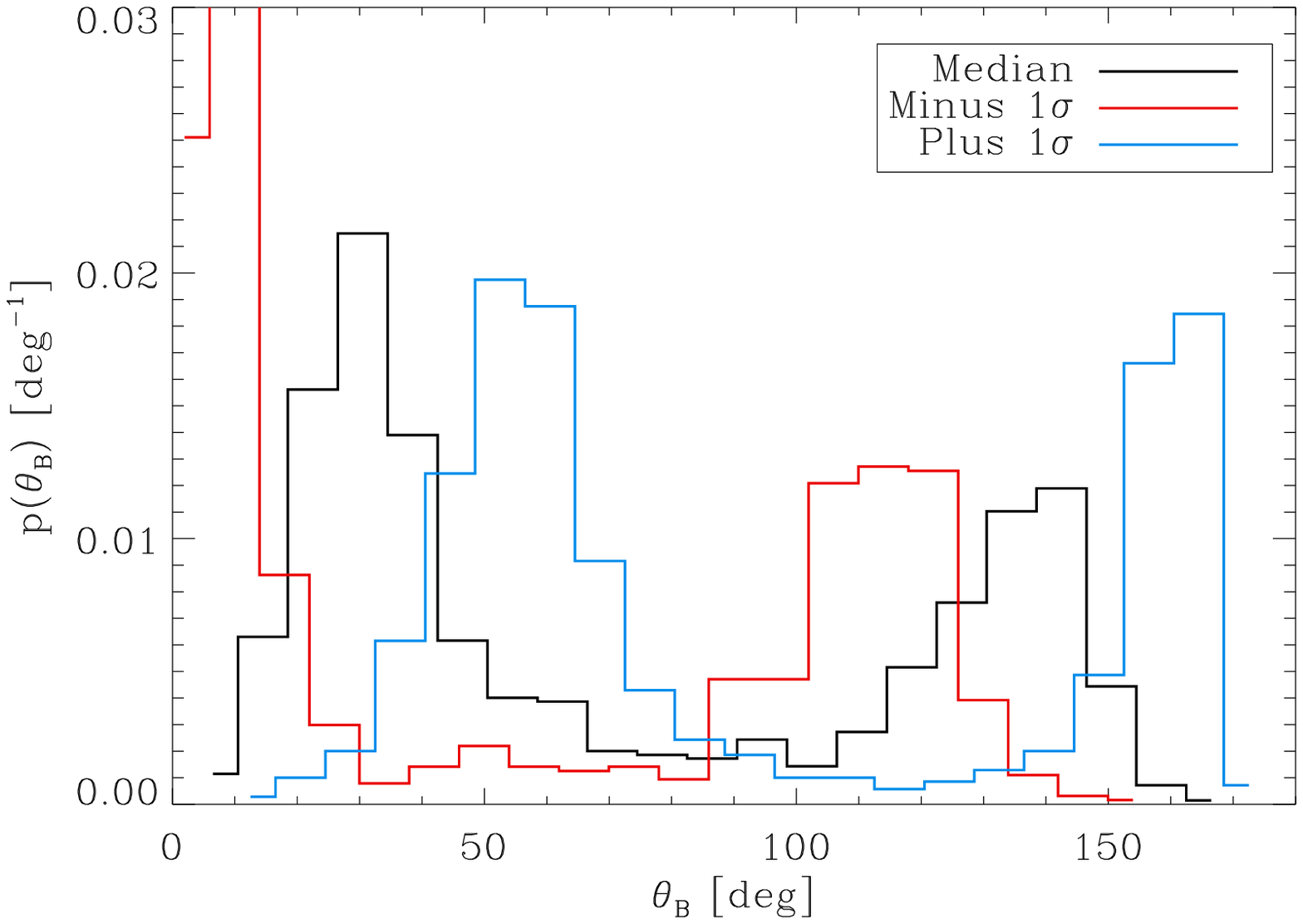}
\includegraphics[width=0.5\textwidth]{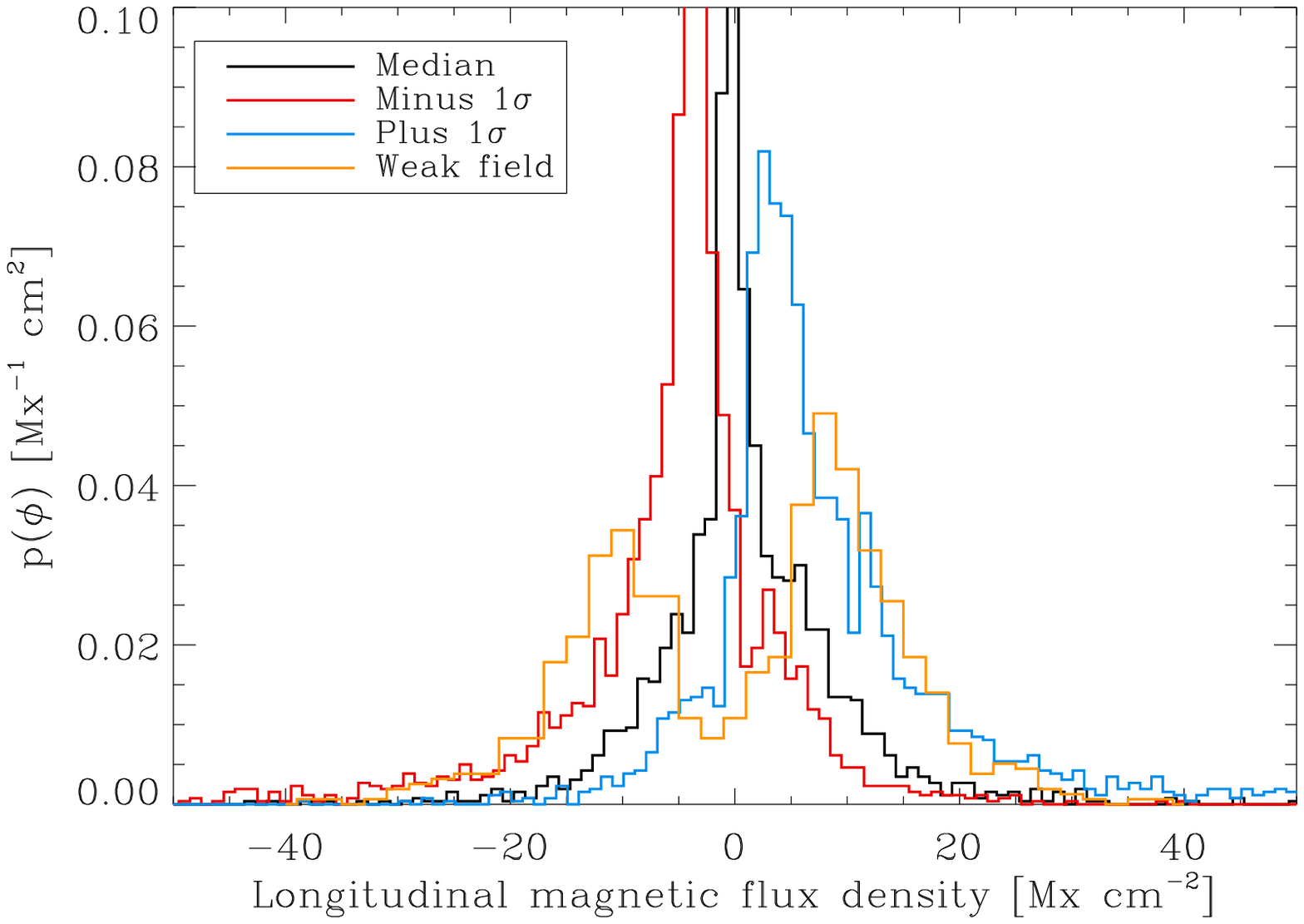}
\caption{Histograms showing the parameters estimated from the marginal posterior distribution
functions for the magnetic filling factor (upper left panel), upper limit of the
magnetic field strength (upper right panel), inclination (lower left panel) and
longitudinal magnetic flux density (lower right panel) for those pixels containing enough information.
Note that this include 100\% of the pixels for $1-\alpha$ and $\phi$, 36.9\% of the pixels for $B$ and only
33.7\% of the points for $\theta_B$. For the magnetic field strength, the results have been separated into granules and intergranules.}
\label{fig:estimation_histograms}
\end{figure*}

After the claims raised by \cite{lites08} and \cite{orozco_hinode07} about the 
large amount of very inclined magnetic fields (with a distribution that peaks
near 90$^\circ$), it is interesting to analyze how much information is still encoded in the Stokes
profiles of the \emph{Hinode} observations under study as analyzed with the
suggested Milne-Eddington model. It is important to point
out that, although the presence of a large amount of inclined fields was inferred
by \cite{lites08} and \cite{orozco_hinode07} with data with the same noise level as those presented
here, it was also confirmed by \cite{lites08} with observations with a noise level a factor $\sim 4$ smaller. The 
middle left panel of Fig. \ref{fig:posterior_examples_kl}
shows the marginal posterior probability distribution functions for the field inclination.
We have verified that it is fundamental to have $D_\mathrm{KL}(\theta_B) \geq 0.5$ in order to end up with a
posterior distribution function clearly different from the prior distribution (dashed line). According
to Table \ref{tab:kl_divergences} and Fig. \ref{fig:kl_histogram}, the number of points fulfilling
this condition is close to 50\%. We have to point out that many of these points
present posterior marginal distributions similar to the green or blue curves in the middle
left panel of Fig. \ref{fig:posterior_examples_kl}. Such posteriors, that discard half of
the range of inclinations but give equal probability to the other half, are a consequence of the 
small Stokes $V$ signal that is still enough to univoquely select the direction of the field
but not its inclination. In other words, the polarity of the field is easily recovered but it
remains impossible to constrain more the inclination. This means that, even for pixels with
low signals (well below the threshold), there is enough information to, at least, detect 
the polarity of the field. Obviously, when the filling fraction of the magnetic component is very small and 
the magnetic field is in the weak field regime, the field cannot be recovered uniquely. This translates
into posterior distribution functions with long tails for the field strength and the field inclination,
as a consequence of the marginalization over the rest of variables \citep[see][]{asensio_martinez_rubino07}.


Concerning the field azimuth (middle right panel of Fig. \ref{fig:posterior_examples_kl}), 
the results are not very encouraging. The majority
of marginal posterior distributions commensurate with the prior distribution. Only in 
approximately 1\% of the field-of-view we find marginal posteriors with a clear peak favoring a certain
subset of azimuths (those corresponding to the white patches of the middle right panel 
of Fig. \ref{fig:kl_images}), while in 
$\sim$4\% of the points, the information encoded in the Stokes profiles allows to roughly determine the
azimuth but with a very large error bar.

The lower two panels of Fig. \ref{fig:posterior_examples_kl} present several posterior
distributions for $B_\parallel$ and $\phi$ for pixels with different values of the 
Kullback-Leibler divergence. The dashed lines show the prior distributions, whose
functional forms are, assuming uniform priors for $\alpha$, $B$ and $\theta_B$:
\begin{equation}
p(B_\parallel) = \frac{1}{1500 \pi} \left\{ -\ln |B_\parallel| + \ln \left[ 1500 + \sqrt{1500^2 - B_\parallel^2} \right] \right\},
\end{equation}
for the longitudinal component of the field, and
\begin{equation}
p(\phi) = \frac{1}{1500 \pi} \int_{|\phi|}^{1500} \frac{2 \ln 2 + \ln 3 + 3 \ln 5 - \ln x}{\sqrt{x^2-\phi^2}} \mathrm{d}x,
\end{equation}
for the longitudinal magnetic flux density. Note that $p(B_\parallel=0)$ and $p(\phi=0)$ tend to infinity in this case
although the distribution has finite area. We believe it is important to point out that
the product of random variables distributed uniformly 
can result in highly non-trivial distributions. Consequently, \emph{it is extremely important to understand which
is the inherent prior information introduced in any inversion method because, when the data is non-informative,
the parameters recovered can be strongly influenced by the prior}.

The posterior distributions for $B_\parallel$ for small values of
$D_\mathrm{KL}(B_\parallel)$ resemble the prior distribution. They become much more distinct when enough
information is encoded in the Stokes profiles, thus resulting in peaked posterior distributions without
extended tails. Interestingly, as a consequence of the multiplication by the well-determined magnetic filling factor,
all the posterior distributions for the longitudinal flux density are quite dissimilar to the prior, except
perhaps those points with small values of the Kullback-Leibler divergence, i.e., $D_\mathrm{KL}(\phi) \approx 1.7$. 

\subsection{Parameter estimation}
\label{sec:parameter_estimation}
According to the results of the previous section, the magnetic filling factor and the upper limit
of the magnetic field strength can be nicely
inferred for a relatively large amount of pixels. Likewise, it is possible to set
constraints on the field inclination for a subset of the field-of-view, while almost no 
pixels contain information about the field azimuth. Since the posterior 
distributions for $1-\alpha$ are fairly peaked, we summarize them using the 
median value, together with the upper 95\% confidence
limits. It is calculated as the value that encloses
95\% of the total area of the posterior distribution starting from $1-\alpha=0$. The results are shown in the
upper left panel of Fig. \ref{fig:estimation_histograms}, where we plot the histogram of
median values in black and the histogram of the $2\sigma$ limit in red, respectively.
The vertical dashed line indicates the position of the most abundant value found by \cite{orozco_hinode07}.
Note that if the plot is
done in linear scale almost all the points collapse to $1-\alpha \approx 0$. Concerning
the confidence intervals, the histograms indicate that they are relatively narrow, completely
discarding magnetic filling factors with $1-\alpha > 0.4$ at 95\% confidence. This upper limit
is in rough agreement with the results presented by \cite{orozco_hinode07}.

Because the shape of the posterior distributions for the magnetic field strength only allows 
us to put upper limits in this parameter, we show in the upper right panel of Fig. 
\ref{fig:estimation_histograms} the
histograms of the horizontal variation of the 68\% and 95\% upper limits of the distributions only for those pixels
having $D_\mathrm{KL}(B) > 1$ (roughly one third of the field-of-view, according to the results shown in
Table \ref{tab:kl_divergences}). For illustration, we have also divided the field-of-view
into pixels belonging to granules and to intergranular lanes. The selection criterion we have used is very simplistic
and it is based on selecting pixels with $I_c/\langle I_c \rangle > 1.02$ as granules and
pixels with $I_c/\langle I_c \rangle < 0.98$ as intergranules. The histograms indicate a
quite unclear difference between the limit field in ``granules'' and ``intergranules'', with
the histogram of the 68\% confidence limit field around 200 G for both cases. The histograms are obviously 
shifted towards higher field strengths for the 95\% confidence histograms but the results
undeniably discard fields above $\sim$900 G. According to the information available in the
Stokes profiles observed with \emph{Hinode}, these histograms are the only statistically relevant 
result that one can give about the magnetic field strength, except for some
individual points where the posterior distribution presents a clear Gaussian-like shape. It is important
to note that these results do not discard 
probability distribution functions peaking at much smaller fields like those present in MHD
simulations \citep{vogler05}, or those inferred from infrared data 
\citep[e.g.,][]{lin95,khomenko03,martinez_gonzalez_spw4_06,marian08} or
lines with hyperfine structure \citep{ramirez_velez08}. However, they discard
pixels in the field-of-view having fields above 900 G. In this sense, the results 
obtained by \cite{orozco_hinode07} using standard Levenberg-Marquardt inversion
schemes are inside the limits posed by our Bayesian analysis.

Only around 50\% of the Stokes profiles observed in the field-of-view contain enough 
information for introducing constraints on the inclination ($D_\mathrm{KL}(\theta_B) > 0.5$). If we
summarize the inferred inclination by the median value of the posterior distribution (without taking
into account its spread), the result is shown in the lower left panel of Fig. 
\ref{fig:estimation_histograms}. Because of the typically extended shape of the posterior
distributions for the field inclination, the results are therefore of reduced interest as representative of the 
statistical properties of the magnetic field inclination in the whole field-of-view.
The histograms of 68\% confidence intervals are also presented. The results indicate that the median
values peak close to 30$^\circ$ and 140$^\circ$ with a dispersion around
these values. This is a consequence of the fact that for many points we can only distinguish
the polarity of the field (with the posterior distribution resembling, for instance, the curve for 
$D_{KL}(\theta_B)=1.1$ in the middle left panel of Fig. \ref{fig:posterior_examples_kl})
and the median value is close to the center of the $[0^\circ,90^\circ]$
or $[90^\circ,180^\circ]$ intervals.

Finally, the lower right panel of Fig. \ref{fig:estimation_histograms} represents the
median (in black line) and the $\pm 1\sigma$ values (red and blue lines) of the magnetic flux density
as obtained from the marginal posterior distributions of the lower right panel of Fig. 
\ref{fig:posterior_examples_kl}. For comparison, we also show the histogram of the
magnetic flux density obtained under the assumption of the weak field regime for those
pixels with signals above 4.5 times the noise level. The magnetic flux density histogram 
estimated under the assumption of
weak field regime lies inside the Bayesian $\pm 1 \sigma$ confidence region. This is a
consequence of the fact that, even if the cross-talk between the strength 
and inclination of the magnetic field vector and the magnetic filling factor cannot be 
disentangled, the product $\phi=(1-\alpha) B \cos \theta_B$ can be estimated with confidence.

\begin{figure}
\plotone{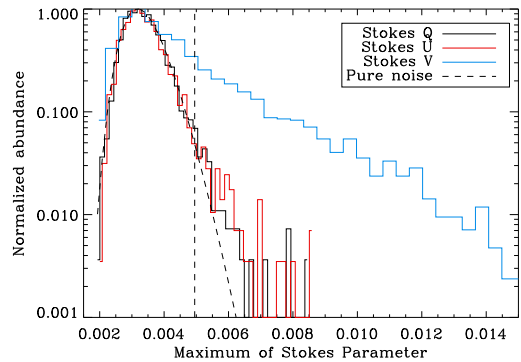}
\caption{This figure shows the histogram of maximum amplitudes of Stokes $Q$, $U$ and $V$. The histogram
in dashed line is the corresponding to pure noise calculated using order statistics. The vertical
dashed line indicates the threshold of 4.5 times the noise level.}
\label{fig:noise_level}
\end{figure}

\begin{figure}[!b]
\plotone{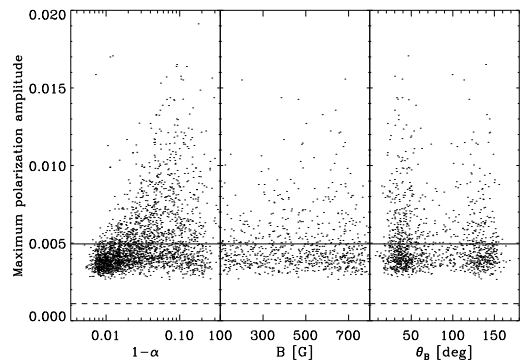}
\caption{Inferred magnetic filling factor, the upper limit of the field and the inclination of the
field versus the maximum amplitude of the polarization profiles, i.e., $\max\{Q(\lambda)/I_c,U(\lambda)/I_c,V(\lambda)/I_c\}$. 
The dashed horizontal line shows the noise level and the solid line presents the threshold of 4.5
times the noise level. This shows that pixels with
amplitudes below and above the threshold of 4.5 times the noise level present similar
behaviors.}
\label{fig:noise_level_pars}
\end{figure}

\subsection{Available information}
\label{sec:information}

The fundamental point to clarify at the light of the results shown in the previous sections
is which is the mechanism (or combination of mechanisms) that is
destroying the information about the magnetic field vector in the observed Stokes 
profiles. To this end, we first analyze the fraction of pixels with signal
above the noise level. Figure \ref{fig:noise_level} shows 
the histogram of the maximum amplitude of Stokes $Q$, $U$ and $V$ in each pixel 
\cite[the quantity chosen by][to set their threshold]{orozco_hinode07}. The histogram
in dashed line presents what one would obtain for a pure noisy case with the
same standard deviation of the \emph{Hinode} data, while the vertical dashed
line indicates the threshold of 4.5 times the noise level. Note that the probability distribution
function of a random variable that results from taking the maximum of a finite number of normal 
random quantities (the pixels in the wavelength direction of the camera) can be
obtained using techniques of order statistics \citep{david_orderstat81}. For the case
of Gaussian noise and 90 points in wavelength direction, this gives the 
skewed distribution shown in Fig. \ref{fig:noise_level}. We have empirically verified that the distribution
peaks at $\sim 2.5\sigma$, where $\sigma$ is the standard deviation of the noise.

In it clear that Stokes $Q$ and $U$ profiles are mainly dominated by noise in a large fraction
of the pixels, with a tail towards large amplitudes that is slightly heavier than
for the case of pure Gaussian noise. The threshold chosen by \cite{orozco_hinode07}
appropriately isolates these points. On the contrary, the Stokes $V$ amplitudes are
not dominated by noise for amplitudes above the threshold. This means that, in the majority of the
pixels, only the (possibly modified) information encoded in Stokes $I$ and Stokes $V$ remains.
Note that the inference is carried out using the full wavelength variation
of the profile, so that it is certain that, even if the amplitude is close to the
noise level, some information may still remain in the correlation between different wavelength points
along the profile. When the stray-light contamination is large in the pixels, the 
information present in Stokes $V$ is also efficiently masked by the noise because the
circular polarization signal decreases. Consequently, once the stray-light contamination is 
set by adapting the model until fitting Stokes $I$, the upper limit in the field
strength is essentially set by the remaining amplitude of Stokes $V$.

The fact that noise is not the only contributor to the lack of information in the
Stokes profiles can be reinforced thanks to the results presented in 
Fig. \ref{fig:noise_level_pars}. In this figure we plot the median values of the
magnetic filling factor, the 68\% upper limit of the magnetic field strength and the
median value of the inclination of the field for those pixels in which 
$D_\mathrm{KL}(1-\alpha)>1$, $D_\mathrm{KL}(B)>1$ and $D_\mathrm{KL}(\theta_B)>0.5$
versus the value of $\max\{Q(\lambda)/I_c,U(\lambda)/I_c,V(\lambda)/I_c\}$. We also show the noise level as a 
horizontal dashed line and 4.5 times the noise level as a horizontal solid line.
All pixels present significant signal of Stokes $Q$, $U$ or $V$, with this
signal always above $\sim 2\sigma$. Except for the stray-light contamination, the points above and below the
threshold set by \cite{orozco_hinode07} present roughly similar results 
and seem to be independent of the amplitude of the profiles. A different behavior is found for the
stray-light contamination. Profiles with signals below the threshold show systematically very large
stray-light contaminations (with $1-\alpha \sim 0.01$), while pixels above the threshold show
systematically smaller contaminations.

\begin{figure}
\plotone{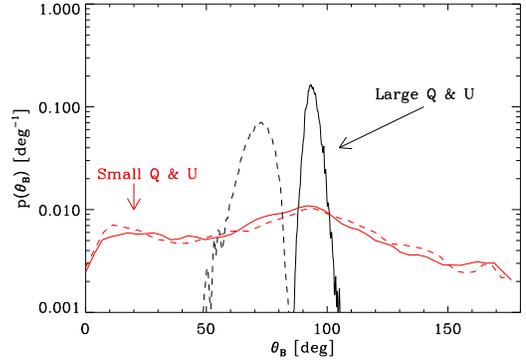}
\caption{Posterior distributions for the field inclination in two pixels
with sizable Stokes $Q$ and $U$ signals (black solid and dashed lines) and in
two pixels with negligible Stokes $Q$ and $U$ signals (red solid and dashed lines).}
\label{fig:posterior_theta_signalQU}
\end{figure}

Although apparently obvious, another quantitative conclusion that we can extract from our 
analysis is that one needs to detect signal in Stokes $Q$ and $U$ in order to get
a good estimation of the inclination. This is demonstrated in the right panel of 
Fig. \ref{fig:posterior_theta_signalQU}. We present the
posterior distribution function for two pixels with negligible Stokes $Q$ and $U$ that
are not detected above the noise level (red curves). The posterior distributions
agree with the prior uniform distribution, so no information about the inclination
is available. On the contrary, for the two pixels for which the Stokes $Q$ and $U$ signal
are well above the noise level the posterior distribution functions are clearly
different from the prior distribution.

\section{Quasi-isotropy of the magnetic field}
\label{sec:isotropy}
\begin{figure*}
\plottwo{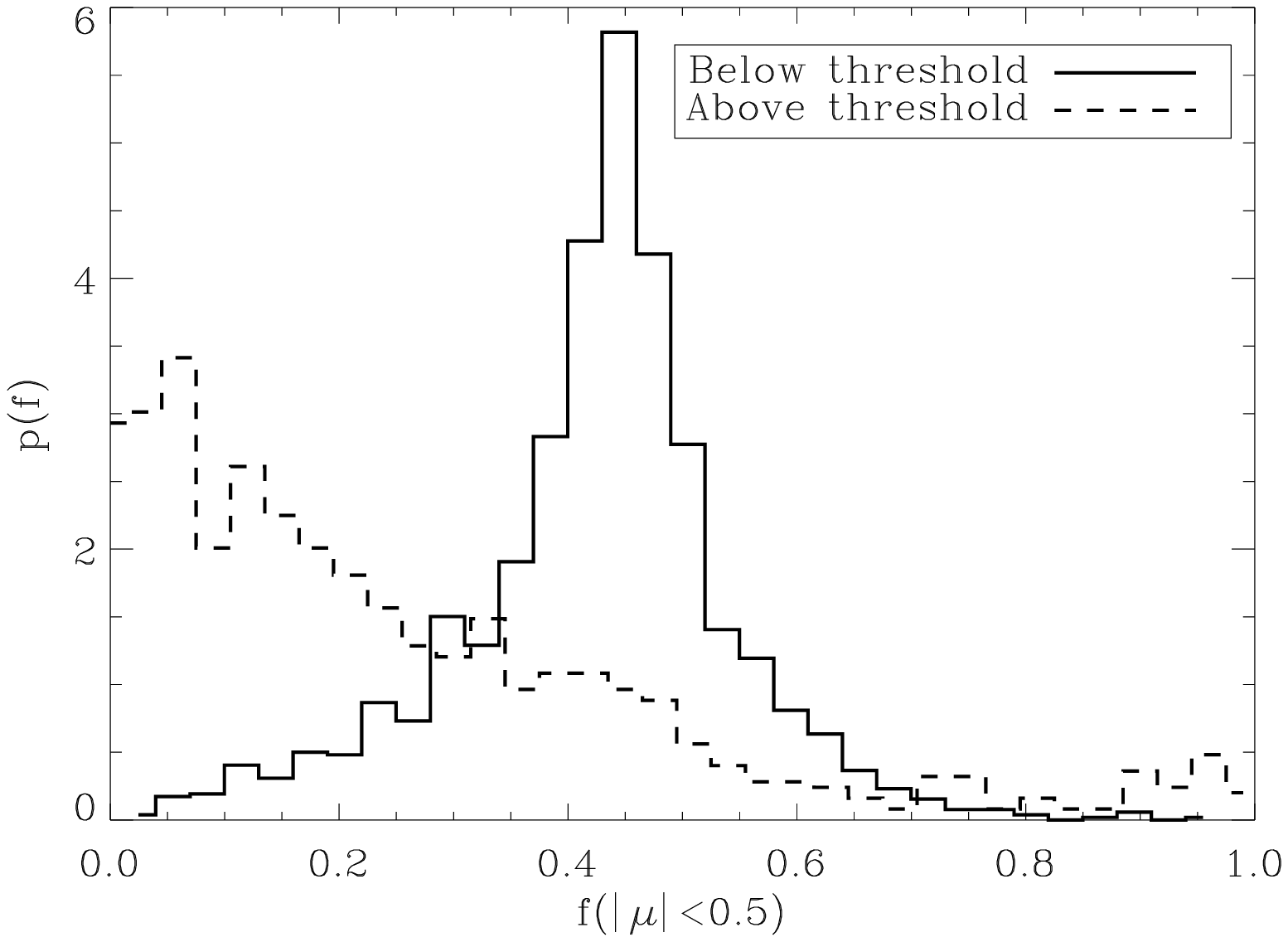}{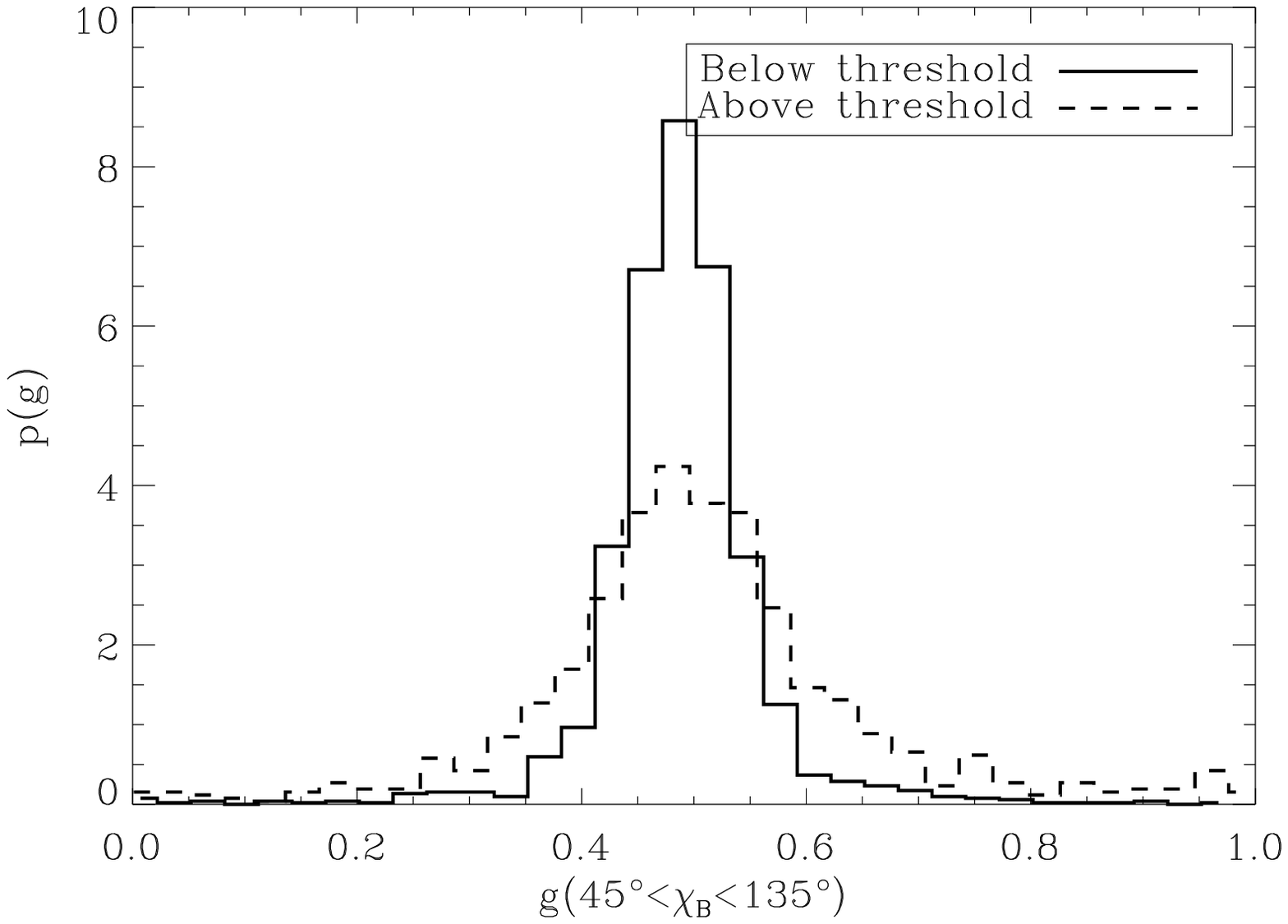}
\caption{Histogram of the posterior distribution mass for the inclination normalized to the
total area of the posterior (left panel) in different inclination regimes. The
right panel shows the posterior distribution mass for the azimuth between 45$^\circ$
and 135$^\circ$. The results suggest that the field is quasi-isotropic for those points whose
polarimetric signal is below 4.5 times the noise level.}
\label{fig:area_inclination}
\end{figure*}

An obvious question that arises after our analysis is whether it is
possible to extract some solid conclusions about the field inclination
from the information available in the observations. In order to answer
to this question, we have calculated, for each pixel, the following quantities, that give the fraction of the total area 
of the posterior distribution function that is enclosed in a given interval:
\begin{equation}
f(\theta_1,\theta_2) = {\int_{\theta_1}^{\theta_2} p(\theta_B) \mathrm{d}\theta_B}, \qquad  
g(\chi_1,\chi_2) = {\int_{\chi_1}^{\chi_2} p(\chi_B) \mathrm{d}\chi_B}. 
\label{eq:prob_inclinations}
\end{equation}
The posterior marginal distributions $p(\theta_B)$ and $p(\chi_B)$ are normalized to unit area.
The quantity $f$ represents the probability that the inclination of the field lies between
$\theta_1$ and $\theta_2$. In an isotropic field, one finds the same amount of vectors whose
inclination lies between 0$^\circ$ and 60$^\circ$ (or between 120$^\circ$ and 180$^\circ$) than between 
60$^\circ$ and 90$^\circ$ (or between 90$^\circ$ and 120$^\circ$). In other words, using Eq. (\ref{eq:prob_inclinations}),
$f(|\mu| <0.5)=f(|\mu| \geq 0.5)=1/2$, with $\mu=\cos \theta_B$. Correspondingly, the quantity $g$ represents the probability that 
the azimuth of the field lies between $\chi_1$ and $\chi_2$. In an isotropic field, one finds the
same amount of vectors with azimuths between 0$^\circ$ and 45$^\circ$ than between 45$^\circ$ and
90$^\circ$ (the same happens for azimuths between 90$^\circ$ and 180$^\circ$). According to 
our imposed prior on the azimuth,
we are not able to distinguish a purely random azimuth from a situation in which we have a non-constrained azimuth.
The left panel of figure \ref{fig:area_inclination} 
shows the histogram of the quantity $f(|\mu| < 0.5)$ for those points whose polarimetric signal is
above the threshold (dashed line) and below the threshold (solid line). Note that, in an isotropic field, the histogram
of $f(|\mu| < 0.5)$ should peak at 1/2 (with some natural dispersion). Although the histogram of pixels
with signals below the threshold (dashed line) clearly point towards a quasi-isotropic distribution,
the histogram of pixels with signals above the threshold (solid line) indicates a clearly non-isotropic distribution.
In this last case, the histogram of high-inclination fields (fields with $|\mu| < 0.5$) peak at values conspicuously
smaller than 1/2. For completitude, the histogram for low-inclination fields (fields with $|\mu| \geq 0.5$) peaks
therefore at values larger than 1/2. Note that,
according to our imposed prior (uniform in $\theta_B$), in the complete absence of 
information on the Stokes profiles, $f(|\mu| < 0.5)$ should
converge to the value $1/3$. The fact that the histogram peaks close to $1/2$ means
that information about the quasi-isotropicity of the magnetic field vector is
still available. Concerning
the azimuth, we plot in the right panel of Fig. \ref{fig:area_inclination} the
histogram of the pixel-to-pixel variation of the posterior mass for the azimuth
between 45$^\circ$ and 135$^\circ$. According to the previous discussion, this should
be close to 1/2 for an isotropic field, although this should also be the case for
completely non-constrained azimuths. We find that this is the case in general for all points
in the field-of-view, irrespective of the amplitude of the polarimetric signal, thus
suggesting the idea that the azimuth of the field is not clearly organized.
Therefore, for pixels with polarimetric signal below 4.5 times the
noise level, the field is apparently quasi-isotropic (as a vector field). We point
out that this result is in agreement with the quasi-isotropic character of the
field found by \cite{marian_clv08}. A more deep 
analysis of improved data is mandatory to evaluate this property of the field.

\section{Comparison with $\chi^2$-minimization}
The model we propose for explaining the observations is exactly that used by
\cite{orozco_hinode07}. Since we are using uniform priors for all the variables and we
are sampling the full posterior distribution, we are also capable of recovering their results 
by just looking at the combination of parameters that gives the maximum value of $p(\thetabold|D,I)$. 
Such model is also the global minimum of the $\chi^2$ merit function and should be equivalent to
that found by \cite{orozco_hinode07} using Levenberg-Marquardt techniques if the method
was successful in locating the global minimum. We point out that
the maximum-a-posterior (MAP) model, although it represents the model that better fits the observations,
is statistically equivalent to all those models inside the 68\% confidence region, as stated above.
Consequently, the presence of noise induces that the MAP model cannot be preferred over the 
rest of compatible models based on statistical justifications.

\begin{figure*}
\plottwo{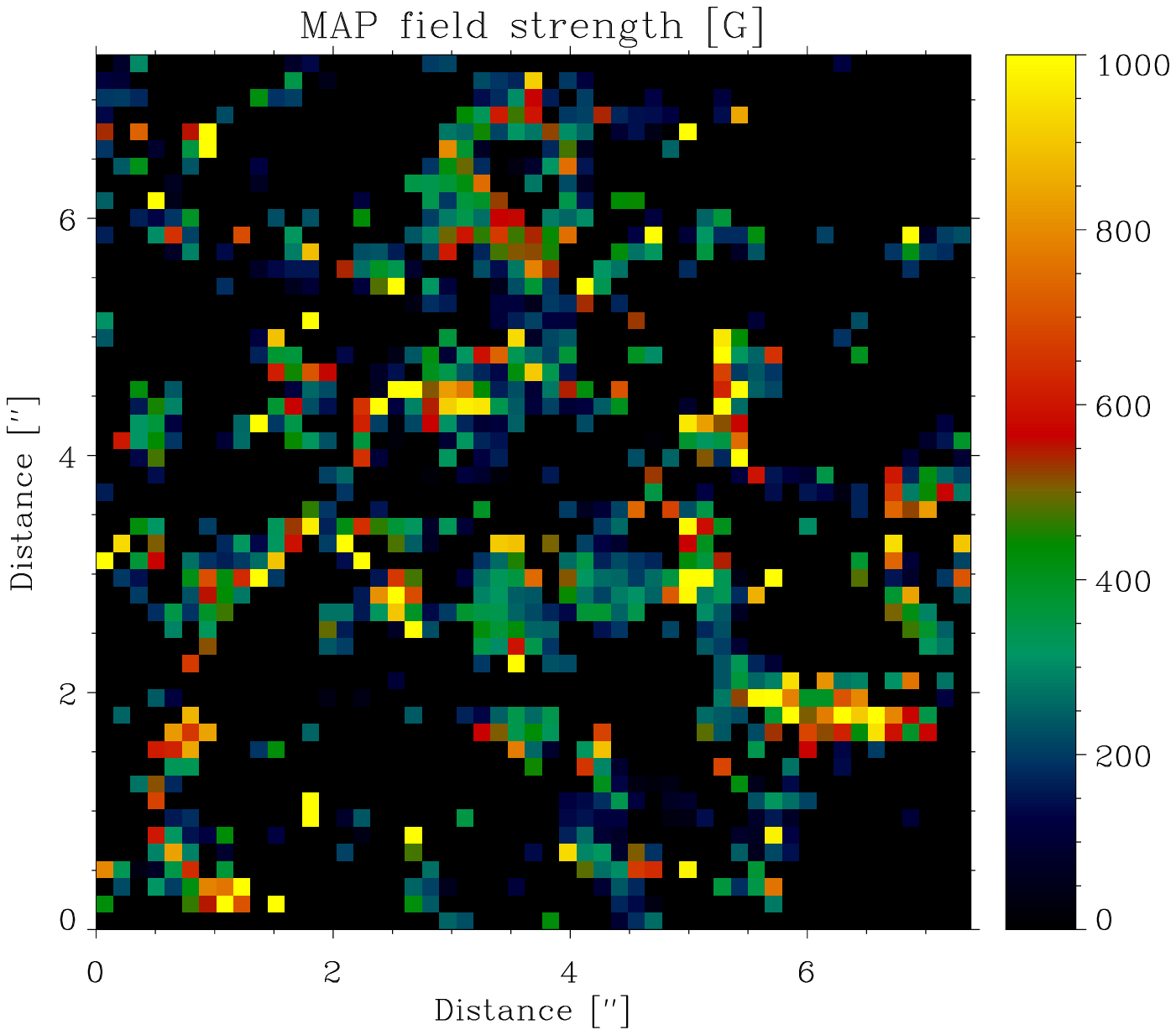}{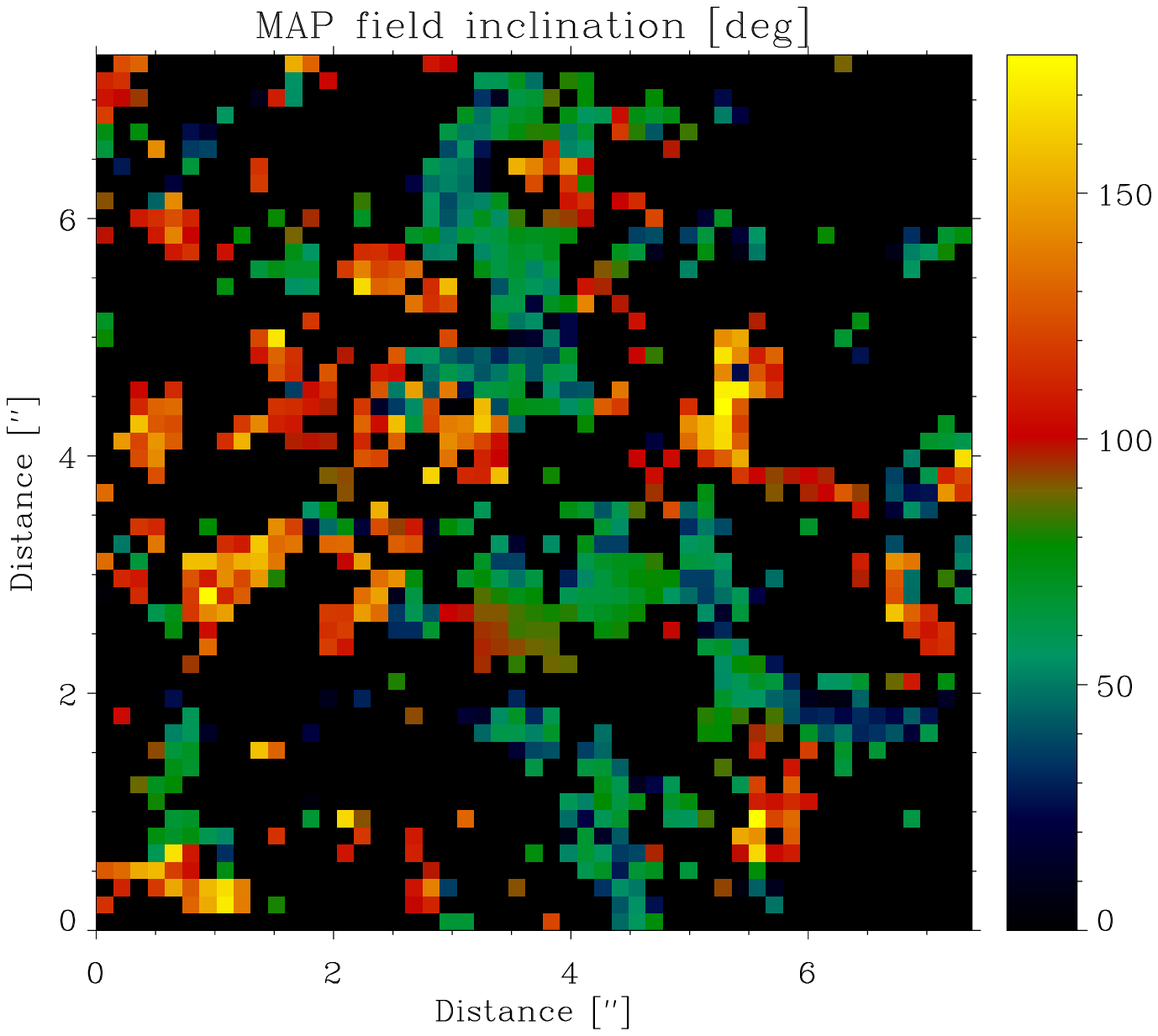}
\caption{Maximum a-posteriori values for the field strength and the field inclination. These results
agree well with those in Fig. 2 of \cite{orozco_hinode07}. Only the result for pixels with polarimetric
signal above the threshold are shown.}
\label{fig:MAP}
\end{figure*}

The correct Bayesian way of giving constraints for individual parameters is done after 
marginalizing the rest of parameters (which are accounted for with their associated probabilities). Unless
the multidimensional posterior distribution is peaked (Gaussian-like, for instance), the MAP
model and the marginal value for each parameter do not coincide. This is especially relevant
when parameters are highly degenerate, which is the case with this dataset.
We have verified that the MAP results agree well with those presented in Fig. 2 of \cite{orozco_hinode07},
as shown in Fig. \ref{fig:MAP}, but they usually differ from the value at the peak of the
marginal distribution. This is a clear indication of the presence of strong degeneracies. 
All points whose polarimetric signal is not above 4.5 times the
noise level are set to black in Fig. \ref{fig:MAP}. The histogram of MAP field inclinations presents
a shape very similar to that found in the lower left panel of Fig. \ref{fig:estimation_histograms},
with two peaks symmetrically placed around 90$^\circ$. Interestingly, the MAP estimation of the
field inclination for those pixels with amplitude signals below the threshold tends to be
very close to 90$^\circ$. Indeed, these pixels also correspond to the points with $D_{KL}(\theta_B) <0.5$.
This suggests that many of the points giving $\theta_B \sim 90^\circ$ in \cite{orozco_hinode07} should
be reconsidered within a Bayesian framework in order to investigate whether this value is really representative
or they are just a subproduct of the lack of information on the Stokes profiles. The fact that
many of these points with inclination angles of the magnetic field close to 90$^\circ$ are located inside regions with
low polarization amplitudes and/or surrounding regions of large signal \citep[see Fig. 1 of][]{orozco_hinode07} points
towards the second possibility. This analysis is left for the future.

\section{Conclusions}
We have applied the \B\ inference code to \emph{Hinode} spectropolarimetric 
observations of the internetwork quiet Sun. The deep analysis carried out under
the Bayesian inference approach and using a simplified Milne-Eddington model of
one magnetic component with a stray-light contamination leads us to conclude that there 
is information in the observables to put upper limits to the magnetic field strength
and to give a good estimation of the stray-light contamination. 
The results indicate that the magnetic field strength is clearly in the hG regime with 95\% confidence,
so that fields in the kG regime are effectively discarded. All the results
presented in this paper depend on the assumption that the model proposed explains
the observables. It might be the case that a simpler or more complex model
allows us to extract more information from the observables because it inherently
reduces some of the degeneracies. This is an issue left for future investigation.

Concerning the field inclination, the information present in the \emph{Hinode} Stokes 
profiles used by \cite{lites08} and \cite{orozco_hinode07} is clearly not enough to tightly constrain it. However, since 
after the Bayesian analysis
we have in hands the full posterior distribution for each parameter and each pixel,
we have been able to conclude that the amount of high-inclination fields (with 
$\theta_B < 60^\circ$ and $\theta_B >120^\circ$) and low-inclination fields
(with $60^\circ < \theta_B < 120^\circ$) is essentially the same for pixels with polarimetric
amplitudes below 4.5 times the noise level. This suggests that
the magnetic fields of part of the quiet solar photosphere are well described by
an quasi-isotropic distribution. This supports the results of \cite{marian_clv08}.
For those pixels with polarimetric amplitudes above the threshold, the inclinations
are clearly constrained.

This paper also demonstrates that, 
although \B\footnote{\B\ is freely available from the author.} is computationally more intensive than
standard Levenberg-Marquardt codes, it is feasible to analyze real data under the
Bayesian framework. The analysis of the $\sim2500$ pixels of the map was done in
less than 8 hours in a standard desktop computer, at a rate of $\sim$10 seconds per pixel.
In spite of the increase in computing time, the amount of information inferred from the analysis is far more complete than that from
standard gradient descent inversions. This has allowed us to verify which parameters are constrained
by the observations and permits the user to yield conclusions based purely on
the observations, clearly showing when conclusions can be flawed by a-priori
assumptions.

\begin{acknowledgements}
We thank L. Bellot Rubio, J. C. del Toro Iniesta, A. L\'opez Ariste, D. Orozco Su\'arez and J. Trujillo Bueno for fruitful discussions.
Financial support by the Spanish Ministry of Education and Science through project AYA2007-63881 is gratefully acknowledged.
\end{acknowledgements}

\end{document}